\documentclass[a4paper,fleqn,usenatbib,useAMS]{mnras}
\usepackage[T1]{fontenc}
\usepackage{ae,aecompl}
\usepackage{graphicx}   
\usepackage{amsmath}    
\usepackage{amssymb}    
\usepackage{hhline}
\usepackage{booktabs}
\usepackage{array}
\usepackage{xcolor}
\usepackage[normalem]{ulem}

\graphicspath{ {Figures/} }

\newcommand\fb{f_{\rm bound}}
\newcommand\Qi{Q_{\rm i}}
\newcommand\Qf{Q_{\rm f}}
\newcommand\Qa{Q_{\rm a}}

\newcommand\Msun{\rm{~M}_{\odot}}

\newcommand\MSun{\rm{~M}_{\odot}}

\newcommand\Rh{R_{\rm h}}

\newcommand\Mgas{M_{\rm gas}}

\newcommand\C{\cal C}
\newcommand\omegab{\Omega_{*,1}}
\newcommand\omegaa{\Omega_{*,2}}

\newcommand\lsf{{\rm LSF}}
\newcommand\vesc{v_{\rm esc}}
\newcommand\erf{{\rm erf}}
\newcommand{\mean}[1]{\left\langle#1\right\rangle}

\title[Gas expulsion in highly substructured embedded star
clusters]{Gas expulsion in highly substructured embedded star clusters} 

\author[J.P.~Farias et
al.]{J.P.~Farias$^{1,2}$\thanks{E-mail:juan.farias@chalmers.se},
  M. Fellhauer$^1$, R. Smith$^3$, R. Dominguez$^1$, J. Dabringhausen$^4$ \\ 
  $^1$Departamento de Astronomia, Universidad de Concepcion, Casilla
  160-C, Concepcion, Chile\\  
  $^2$Department of Space, Earth and Environment, Chalmers University of Technology,
  Onsala Space Observatory, \\ ~Observatoriev{\"a}gen 90, 43992 Onsala, Sweden \\ 
  $^3$Korea Astronomy and Space Science Institute, 766, Daedeokdae-ro, Yuseon-gu, Daejeon, 34055, South Korea \\ 
 $^4$Charles University, Faculty of Mathematics and Physics, Astronomical
 Institute, V Hole\v{s}ovi\v{c}k\'ach 2,
 CZ-180 00 Praha 8,\\ ~Czech Republic}

\date{Accepted -----. Received -----; in original form -----}

\pubyear{2017}

\begin{document}
\label{firstpage}
\pagerange{\pageref{firstpage}--\pageref{lastpage}}
\maketitle

\begin{abstract}
We investigate the response of initially substructured, young, embedded star clusters to
instantaneous gas expulsion of their natal gas.  We introduce primordial substructure to
the stars and the gas by simplistically modelling the star formation process so as to
obtain a variety of substructure distributed within our modelled star forming regions. We
show that, by measuring the virial ratio of the stars alone (disregarding the gas
completely), we can estimate how much mass a star cluster will retain after gas expulsion
to within 10$\%$ accuracy, no matter how complex the background structure of the gas is,
and we present a simple analytical recipe describing this behaviour.  We show that the
evolution of the star cluster while still embedded in the natal gas, and the behavior of
the gas before being expelled, are crucial processes that affect the timescale on which
the cluster can evolve into a virialized spherical system.  Embedded star clusters that
have high levels of substructure are subvirial for longer times, enabling them to survive
gas expulsion better than a virialized and spherical system.  By using a more realistic
treatment for the background gas than our previous studies, we find it very difficult to
destroy the young clusters with instantaneous gas expulsion. We conclude that gas removal
may not be the main culprit for the dissolution of young star clusters. 
\end{abstract}



\section{Introduction}
\label{sec:intro}

The vast majority of stars appear to form in groups from dozens to thousands of members
inside molecular clouds \citep{Lada2003, Bressert2010, King2012}.  However, embedded star
clusters do not hold their natal gas for long.  Even before forming low-mass stars that
reach the main sequence, proto-stars already inject energy into the surroundings gas via
proto-stellar jets, and when a massive star forms, large amounts of energy are radiated
into the field.  Finally, the first supernovae explodes and, depending of the size of the
region, could remove any remaining gas in the cluster \citep[see][]{Lada2003}.  Star
formation is observed to be a highly inefficient process.  It is estimated that at most
30\% of the gas ends up converted into stars \citep{Dobbs2014, Padoan2014} thus it has
been argued that the gas removal process is highly destructive and can disperse most of
the star clusters into the field \citep[e.g.][]{Hills1980, Elmegreen1983,
Verschueren1989}. 

Several authors have examined effects of gas loss in star clusters
\citep[see][]{Tutukov1978, Hills1980, Mathieu1983, Elmegreen1983, Lada1984,
Elmegreen1985, Pinto1987, Verschueren1989, Goodwin1997, Goodwin1997a, Geyer2001,
Boily2003, Boily2003a, Goodwin2006, Bastian2006, Baumgardt2007, Parmentier2008,
Goodwin2009}, but most of these works have concentrated on gas loss from clusters in
which the stars and gas are both dynamically relaxed and in global virial equilibrium
identifying the global star formation efficiency (SFE) and the gas expulsion rate as the
parameters that decide how star clusters respond to gas expulsion.  But, star clusters
form from hierarchically substructured molecular clouds and stars are born inside that
substructure \citep{Whitmore1999, Johnstone2000,Kirk2007, Schmeja2008, Gutermuth2009,
diFrancesco2010, Koenyves2010, Maury2011, Wright2014}.  Initially substructured clusters
need to relax for at least one crossing time to reach a spherical and virial equilibrium
distribution.  During this relaxation process the global dynamical state of a cluster can
be very different from virial equilibrium \citep[see][]{Smith2011} and depending of the
size of the cluster, stellar feedback can remove the gas well before the star cluster is
completely relaxed.  

\citet{Verschueren1989} and \citet{Goodwin2009} noted that the exact dynamical state of
clusters at the moment of gas expulsion is extremely important and the SFE alone can not
tell what will be the fate of a cluster.  The inclusion of primordial substructure in the
studies \citep{Smith2011,Smith2013} also show that the SFE is not a good estimator even
when the cluster match virial equilibrium velocities, because the SFE is a global and
static parameter that does not account for the expansion and contraction of the cluster
during the relaxation phase. 

\citet{Smith2011} introduced the Local Stellar Fraction (LSF) defined
as:  
\begin{eqnarray}
  {\rm LSF} = \frac{M_*(r<\Rh)}{M_*(r<\Rh) + M_{\rm
      gas}(r<\Rh)} 
\end{eqnarray}
where $\Rh$ is the radius that contains half of the total mass in stars.  $M_*$ and
$M_{\rm gas}$ is the mass of the stars and the gas, respectively, measured within $\Rh$.
It has been shown that the LSF is a much better indicator of cluster survival than
the SFE \citep{Smith2011}.

In our previous work \citep{Farias2015} we have quantified the relevance of the dynamical
state of initially substructured clusters, measured by the pre-gas-expulsion virial ratio
$Q_{\rm f}$, introducing a very simple analytical model that only depends on the LSF and
$Q_{\rm f}$.  This model predicts quite well the amount of stellar mass that remains
bound after gas expulsion even when gas is removed at very early stages of the star
cluster evolution.  Such models were tested utilizing initially substructured
distributions of stars embedded in a static and smooth background potential. The argument
to include primordial substructure for the distribution of the stars is the observational
evidence that star formation follows spatial distribution of the gas, which is
substructured. This substructure is molded by the internal supersonic turbulence in the
gas, while the source and nature of this turbulence is still a matter of debate. To
complete the picture, we give the gas in this paper the ability to evolve and interact
with the stars. We also include primordial substructure in the gas and a consequent
stellar distribution by emulating the star formation process with an \emph{ad hoc} recipe
and expelling the gas instantaneously at different embedded star cluster ages. 

Before testing the analytical model of \citet{Farias2015}, we modify it in order to
account for the different gas and stellar spatial distributions and we explain why the
simplistic model fails at certain ranges of LSF. We use this model to show how the
$\fb$-LSF trend we have found in previous studies depends on the spacial and dynamical
configuration of the stars and the gas. We find that the model might not be accurate for
more exotic configurations, that young embedded star clusters might have. Therefore, we
test this new model in a more realistic scenario and show an alternative to the previous
estimations.

In Section~\ref{sec:models} we describe the modified analytical approach that we use to
predict the outcome of our simulations.  In Section~\ref{sec:ic} we describe the
numerical methods and assumptions used in the star formation simulations.  We show our
results in Section~\ref{sec:results}, and we discuss and present our conclusions in
Section~\ref{sec:conclusion}.

\section{Analytical approach}
\label{sec:models}

In \citet{Farias2015} we introduce a very simple model that works fairly well in
predicting the amount of bound mass that clusters can retain after instantaneous gas
expulsion.  In this model, we made several assumptions that may not hold in realistic
clusters.  One important assumption was that stars and gas follow approximately the same
distribution. Thus, we expressed the potential energy on the cluster before gas expulsion
as: 
\begin{eqnarray}
  \omegab &\sim& -M_*\frac{GM_{\rm tot}}{\Rh}
\end{eqnarray}
where $M_{\rm tot}$ is the total mass in stars and gas in the cluster.  This assumption
could be particularly important in substructured embedded star clusters.  Even though we
expect that stars and gas follow a similar distribution initially, stars decouple very
fast from the gas and form their own hierarchy.  This happens because stars and gas
respond to very different physical mechanisms \citep{Girichidis2012}.  In this section we
reconstruct the \citet{Farias2015} analytical model in a more general way and provide an
alternative method to estimate the bound fraction based only on the properties of the
stellar distribution. 

\subsection{Estimating the final bound fractions}
\label{sec:model1}

We consider an arbitrary distribution of stars embedded in an arbitrary gas distribution.
The gas and star distributions are not necessarily spherical or, indeed, similar to each
other at the exact moment when instantaneous gas expulsion begins.  We assume that the
stellar distribution follows a Maxwell-Boltzmann velocity distribution.  We will denote
quantities just before gas expulsion with subscript 1 and just after gas expulsion with
subscript 2.  Considering the different spatial distributions, the potential energy of
the stars just before gas expulsion is given by: \begin{eqnarray}
  \label{eq:omega1}
  \omegab &=& - A\frac{GM_*^2}{\Rh} - B\frac{GM_*\Mgas}{\Rh}
\end{eqnarray}
where we use the same scale radius in both contributions.  In this work we choose $\Rh$
to be the half mass radius of the stellar cluster.  $A$ and $B$ are structural parameters
that depend on the distributions of the stars and the gas, as well as the chosen scale
radius $\Rh$.  $A$ only depends on the stellar component while $B$ is more complicated,
depending on how the stellar component is distributed with respect to the gas
distribution.

Thus, the parameters $A$ and $B$ are basically a measure of the structure of each
potential and tell us about the geometrical distribution of the star clusters. They are
given as 
\begin{eqnarray}
        A &=& - \Omega_{*,*}\frac{R_{\rm h}}{M_*^2}
\end{eqnarray}
and
\begin{eqnarray}
        B &=& - \Omega_{\rm *,gas}\frac{R_{\rm h}}{M_*M_{\rm gas}},
\end{eqnarray}
where $\Omega_{*,*}$ and $\Omega_{\rm *,gas}$ are the potential energy of the stars due
to themselves and due to the gas respectively. In our simulations, we can estimate $A$
and $B$ numerically at any given time, because we have full access to the spatial
3D-distributions of gas and stars. However, given the complicated substructure of the gas
in particular, it would be impossible to estimate these parameters for an observed young
star cluster. The reason is that, observationally, we only have the 2D-projections of the
densities of stars and gas along our line of sight, even in the best case scenario (i.e.
no absorption or saturation).

We can use the LSF to estimate the total mass in the region where the stars are present,
i.e., $M_{\rm tot} = \Mgas +M_* \sim M_*/$LSF.  From here we can obtain the amount of
gas in this region as:
\begin{eqnarray}
  \Mgas &\approx& \frac{1-{\rm LSF}}{\rm LSF} M_*.
\end{eqnarray}

After gas expulsion, the potential energy of the cluster only depends on the stellar
distribution.  Considering instantaneous gas expulsion, stars have no time to change
either their velocities or their positions.  Thus the kinetic energy remains equal, i.e.,
$T_*=T_{*,1}=T_{*,2}$ and the structure parameter $A$ remains the same as well.  Thus the
potential energy after gas expulsion is:
\begin{eqnarray}
  \omegaa &=& -A\frac{GM_*^2}{\Rh}.
\end{eqnarray}
We can rewrite Eq.~\ref{eq:omega1} as
\begin{eqnarray}
  \omegab &=& \frac{\omegaa}{A} \left[ A + \frac{(1-\lsf)}{\lsf} B
  \right]\\ 
  &=& \eta\omegaa,
  \label{eq:omegas}
\end{eqnarray}
where we define
\begin{eqnarray}
  \eta(\lsf,A,B)&=& 1 + \frac{(1-\lsf)}{\lsf}\frac{B}{A}.
  \label{eq:eta}
\end{eqnarray}
The escape velocity after gas expulsion can be expressed by
\begin{eqnarray}
  \vesc&=& \sqrt{-2\frac{\omegaa}{M_*}}.
  \label{eq:vesc1}
\end{eqnarray}

Using the definition of the virial ratio and Eq. ~\ref{eq:omegas},
\begin{eqnarray}
  \Qf &=& \frac{T_*} {-\omegab}\\
  &=& \frac{T_*}{-\eta\omegaa},
  \label{eq:Qf}
\end{eqnarray}
and assuming that the stars follow a Maxwellian velocity distribution, the total kinetic
energy of the stars can be written as
\begin{eqnarray}
  T_*&=& \frac{3\kappa}{2} M_* \sigma_*^2,
\end{eqnarray}
where $\kappa=\pi/(3\pi-8)$. Thus, we can rewrite Eq.~\ref{eq:vesc1} as
\begin{eqnarray}
  \vesc&=& \sqrt{\frac{2T_*}{\eta\Qf M_*}}\\
  &=& \sqrt{\frac{3\kappa}{\eta\Qf}} \sigma_*.
  \label{eq:vesc}
\end{eqnarray}

A reasonably first guess for the bound fraction would be the fraction of stars with
velocities below the escape velocity.  In a Maxwellian velocity distribution this
fraction comes from the Cumulative Density Distribution (CDF) evaluated in $v=\vesc$.
With respect to $\sigma_*$, this function is
\begin{eqnarray}
  F(<X)&=& \erf\left( \frac{1}{\sqrt{2}}X \right) -
  \sqrt{\frac{2}{\pi}}X\exp\left(-\frac{X^2}{2}\right),
\end{eqnarray}
where $X=v/\sqrt{\kappa}\sigma_*$.  Evaluating in $\vesc$ and using
Eq.~\ref{eq:vesc}, we obtain
\begin{eqnarray}
  \fb=\erf\left( \sqrt{\frac{3}{2\eta\Qf}}
  \right)-\sqrt{\frac{6}{\pi\eta\Qf}}\exp\left( 
    -\frac{3}{2\eta\Qf} \right).
  \label{eq:fbound}
\end{eqnarray}
Note that for $B/A = 1$, $\eta=1/\lsf$ and Eq.~\ref{eq:fbound} is then equivalent to the
\citet{Farias2015} model.

\subsection{An alternative approach}
\label{sec:model2}

Using the same model, it is possible to avoid measurements of the $\eta$ function as
described before. Considering the virial ratio of the cluster right after gas expulsion,
\begin{eqnarray}
  Q_{\rm a} &=& -\frac{T_*}{\omegaa},
\end{eqnarray}
and using Eq. \ref{eq:Qf} we obtain that
\begin{eqnarray}
  \eta &=& \frac{\Qa}{\Qf}.
  \label{eq:qratio}
\end{eqnarray}
Therefore, Eq. \ref{eq:fbound} becomes:
\begin{eqnarray}
  \fb=\erf\left( \sqrt{\frac{3}{2\Qa}}
  \right)-\sqrt{\frac{6}{\pi\Qa}}\exp\left( 
    -\frac{3}{2\Qa} \right).
  \label{eq:fbound2}
\end{eqnarray}

We emphasize that $\Qa$ is the virial ratio of the cluster after gas expulsion.  As we
are dealing with instantaneous gas expulsion, it is also the dynamical state of the
cluster right before the gas is expelled, ignoring completely the presence of the gas.
We will show in Section~\ref{sec:results} how well this simplified measure fares in
predicting the results of our simulations.

In this approach, only one parameter is necessary to estimate $\fb$.  It is still
challenging to measure such a value, where the most problematic issue is to estimate
$\omegaa$.  However, this result highlights that the specific geometry of the gas and the
stars is not really important.  What is important is the dynamical state of the cluster
if we suddenly remove the gas.  In general, a system with $Q>1$ is said to be unbound.
According to Eq.~\ref{eq:fbound2} a $\fb$ fraction of the cluster is still bound and the
cluster will not be completely dissolved, e.g., for a cluster with $\Qa=1$ we estimate
that a 60\% of the stars will stay bound.

\section{Initial Conditions and Numerical Methods}
\label{sec:ic}

In \cite{Farias2015} we evolved fractal distributions of stars embedded in a static,
smooth background potential to mimic the gas, which we assume follows a Plummer density
profile. Here, we advance the picture of hierarchical star cluster formation further by
introducing a dynamically live and primordially substructured gas background. With this
addition to our models, we have to change the numerical integrator used in
\citet{Farias2015}, since it is not designed to include an hydrodynamical system like the
gas. 

Before advancing that further, we first wish to test if the inclusion of a live gas
background, and also the use of a different code, might affect the results found in
\citet{Farias2015}.  In particular, we test if there is any change in the $\fb$-LSF trend
in the case of $\Qf=0.5$ for a smooth Plummer background gas.  

We then proceed by setting up two different numerical experiments described in
Section~\ref{sec:setup}.

First, as a control test, we set up the same systems as previously studied in
\citet{Farias2015}, with the only difference that gas is now able to evolve.  

In the second experiment we create substructured initial conditions by evolving a
turbulent uniform sphere of gas in a star formation-like fashion in order to generate
stellar and gas substructure throughout the model star forming region.  Our approach was
to evolve the initially uniform ad turbulent sphere of gas, while applying our own
\emph{ad hoc} star formation prescription. We do not use sink particles as we do not want
to include the effects of star particles with varying masses at this stage. Indeed, the
effects of the inclusion of an initial mass function is being prepared in parallel to
this work by Dominguez et al.\ (in preparation). We are not concerned with implementing
star formation in the most accurate way possible, as such simulations are inevitably very
expensive computationally, and could potentially exhaust all of our resources in just a
single simulation. The reader should be aware that these are not formal star formation
simulations since we can not follow fragmentation correctly, neither use stellar
accretion models (see Section~\ref{sec:icdiscussion}).  However, the end result of
evolving the initially uniform and turbulent spheres of gas, while applying our
numerically cheap star formation prescription, is that we can generate large numbers of
substructured clusters, in which the stars roughly follow the substructure of the gas in
a manner that broadly mimics the substructure in real star forming regions. These
substructured conditions are then used as initial conditions for our gas expulsion tests,
and we have sufficiently large samples of such initial conditions that our results are
statistically valid, and not dominated by cluster-to-cluster variations.

In our experiments, gas is always expelled instantaneously. As such, the resulting bound
fractions can be interpreted as the lower limits of cluster survival, since instantaneous
gas expulsion is the most destructive mode of gas loss 
\citep[see e.g.][]{Baumgardt2007,Smith2013}. 

In Section~\ref{sec:setup} we explain the numerical setup that we use in both experiments
with a live gas background. The details of the smooth background gas simulations are
described in section~\ref{sec:plummersetup}. The initial conditions and details of the
substructured simulations, as well as the ad hoc star formation recipe we use, are
explained in Section~\ref{sec:starformsetup}. 

\subsection{Modeling stars embedded in a live gas}
\label{sec:setup}

We perform simulations utilizing the Astrophysical Multipurpose
Software Environment \textsc{Amuse} \citep{PortegiesZwart2013,
  Pelupessy2013,Mcmillan2012}.  \textsc{Amuse} is a high level interface developed
in \textsc{Python} allowing the user to couple different systems
evolving in different physical domains and scales. In our case those
domains are: Purely gravitational (stars); and a self-gravitating
hydrodynamical fluid (gas). 

The equations of motion for the stars are solved by using the \textsc{Amuse Ph4}
dynamical module  \citep{Mcmillan2012} which is an MPI parallel fourth order Hermite
integrator \citep[see e.g.][]{Makino1992} with block timestep scheme.  The gas is modeled
with the \cite{Springel2002} conservative Smoothed Particle Hydrodynamics (SPH) scheme
implemented by the code \textsc{Fi} \citep[][see also \citealt{Hernquist1989,
Gerritsen1997}]{Pelupessy2004,Pelupessy2005} which uses the \cite{Monaghan1985} kernel
and computes the self-gravity of the gas using the \citet{Barnes1986} tree scheme. We
have adopted viscosity terms $\alpha=0.5$ and $\beta=1$ (half the commonly adopted
values) since we only expect relatively weak shocks caused mainly by gravitational
collapse.  However, we have tested sensitivity of our results to this choice and find it
is of negligible importance, perhaps due to the lack of strong shocks that develop during
our modelling. In our simulations gas and stars interact only by gravity, we do not
include feedback.  Thus we couple both systems using the \textsc{Bridge} scheme
\citep{Fujii2007} that manages the perturbation of one system onto the other by
gravitational velocity kicks in a Leapfrog timestep scheme \citep[see also][for a similar
setup]{Pelupessy2012}.  Interactions between stars and gas are done symmetrically, i.e.\
utilizing the same method to calculate the gravity of the systems in both directions
(stars perturbed by the gas and gas perturbed by the stars). For this, we choose to use
the \citet{Barnes1986} tree scheme. Such a configuration has proven to be most accurate,
with an energy error below 1\% at all times. 

\begin{table*}

\label{tab:setup}
\centering
\caption {Summary of the constraints used in this work to model young
  star clusters from their parent turbulent molecular cloud to the
  final gas free star cluster remnant.  First column shows the physical
  stage modeled by the method, second column shows comments about the
  constraints related to the stellar component of the cluster, third
  row shows whether the Bridge integrator is enabled for the mutual
  interaction between gas and stars and fourth column shows comments
  about constraints and initial conditions related to the gaseous
  component such as the EOS used in the corresponding phase and the
  velocity field used as initial condition.} 
\begin{tabular}{|m{0.6in}||m{1.4in}|m{0.6in}|m{1.4 in}|}\cline{2-4}
 
 \multicolumn{1}{c|}{ } & 
 \center \bf{\Large{Stars}}  & \center \textsc{BRIDGE} & 
 \center \bf{\Large{Gas}} \tabularnewline \hhline{-::===}

 \center \large{Collapse phase}  & \center ----- & \center Off & 
 \vfill \small{EOS :\textbf{ Isothermal}} \par Initial velocity field:\par   
 \centering $P(k)\propto k^{-4}$ \tabularnewline \hhline{-||---}

 \center \large{Star formation phase} & \center \small{1 star $ = \begin{array}{c} 
         N_{\rm smooth}~{\rm bound} \\ {\rm SPH~particles} \end{array}$} 
         & \center On & \small{EOS: \textbf{Isothermal} \par 
         If $h_{\rm i}< h_{\rm crit}$ then: \par \quad check for star formation \par 
         \quad criterion }  \\ \hhline{-||---}

 \center \large{Embedded phase}   & \center \small{1000 equal mass stars \par 
 $m_{\rm star,i}=0.5 \Msun$} & \center On & 
 {\small{EOS: \textbf{Isothermal} \par  Self-gravity : Off} \par \hrule
 \small{EOS: \textbf{Adiabatic}, $\gamma=5/3$ 
 \par Self-gravity: On}}
 \tabularnewline \hhline{-||---}

 \center \large{Gas free phase} & \center \small{Evolution continues for 15 Myr}& 
 \center Off & \center ----- \tabularnewline \hhline{-||---}

\end{tabular}

\end{table*}

\subsection{A smooth live gas background}
\label{sec:plummersetup}

Our first step in advancing the complexity of our simulations of young embedded star
clusters is to change the static background Plummer potential
previously used in \cite{Farias2015} for a live Plummer sphere of gas 
that is affected by the gravity of the stars. 

We take a set of 20 fractal distributions with fractal dimension of
$D=1.6$ \citep[see][]{Goodwin2004} and $N=1000$ equal mass stars with
$M_{\rm i}=0.5$ M$_\odot$ in a radius of 1.5 pc.  These stellar
distributions are embedded in a Plummer sphere of gas of $R_{\rm pl}=1$ pc and $M_{\rm
pl}=3472$ M$_\odot$, ensuring a global ${\rm SFE}=0.2$ inside the radius of the stellar
distribution. 

We use an adiabatic equation of state (EOS) with adiabatic index of $\gamma=5/3$ with no
cooling or heating recipes. The internal energy of the gas is scaled to account for the
extra mass (the stars) inside the sphere, so that initially the gas is in equilibrium and
subsequent perturbations are only caused by the relaxation of the stars. The stellar
velocities are scaled in order to obtain initial virial ratios of $Q_{\rm i}=0.0$ and
0.5. The gas is modeled with $N_{\rm gas}=100$k SPH particles and a neighbour number $N_{\rm
nb}=64$, which is enough to prevent unphysical scattering and to reproduce the structure
of the Plummer sphere \citep[see Appendix \ref{sec:resolution} and
also][]{Hubber2011,Hubber2013} 

The gas is expelled instantaneously at a
specific point in the evolution of the clusters, namely when the virial ratio increases
to $Q_{\rm f}=0.5$ again after the second full oscillation around $Q_{\rm f}=0.5$ since
the start of the simulation, i.e. at the next passing of $Q_{\rm f}=0.5$ after the green
dashed line in Fig.~1 in \citet{Farias2015}.

\subsection{Creating substructured embedded star clusters}
\label{sec:starformsetup}

In order to create initially substructured stellar and gaseous distributions, we evolve a
turbulent sphere of gas with an isothermal equation of state at $T=10$~K, a radius of
$R_{\rm cl}=1.5$~pc, and a total mass of $M_{\rm gas,0} = 2500 \MSun$.  The gas is
modeled utilizing $N_{\rm gas}=250\,\rm k$  SPH particles and $N_{\rm nb}=50$\footnote{We
have decreased $N_{\rm nb}$ from 64, when using a Plummer sphere, to 50 in this set up in
order to force a resolution of 0.5$\MSun$, which is the mass of the stars we are
attempting to form, without compromising performance.}. The cloud is initially perturbed
utilizing a turbulent velocity field with energy injection mainly on the large scales
(see Section~\ref{sec:collapsephase}). We do not continuously drive turbulence in any of
these simulations, beyond the initial conditions.  We evolve the cloud with an ad hoc
star formation recipe (described in Section~\ref{sec:starformationphase}) forming equal
mass particles of $0.5\Msun$ until we match a SFE of 0.2, i.e., 1000 stars.  As a result
we obtain a filamentary cloud of gas and a consequent stellar distribution that we use as
initial condition for further evolution.  For convenience, we choose $t=0$ as soon as the
desired 1000 stars has formed, although the time to reach this stage does vary between
realizations. At this time, we switch the global EOS of the gas from isothermal to
adiabatic (with adiabatic index $\gamma=5/3$) and follow the evolution of the embedded
star cluster. We expel the gas instantaneously at $t=0$, 1 and 2 Myr and follow the gas
free cluster until $t=15$ Myr.

The different stages of star cluster process needs special considerations and methods,
then for the numerical treatment of the embedded star cluster, we split the simulation
into four stages: 
\begin{itemize}
\item {Collapse phase:} Evolution from an initially spherical, uniform, turbulent gas
        cloud until the star formation criteria is first met.
\item {Star formation phase:} Continues until the desired SFE is met.
\item {Embedded phase ($t\equiv0$):} Starts when we switch the global EOS to adiabatic
        and continues until we decide to expel the gas.
\item {Gas free phase:} The stage after gas expulsion where only the
        stars in the cluster evolve until $t=15$~Myr to make sure all escapers are
  far from the main cluster.  
\end{itemize}
We emphasize that the collapse phase and star formation phase  can be considered an
approach to generate a substructured distribution of stars and gas, that is then used as
initial conditions for our numerical experiment, during the embedded and gas free phase.
We summarize the numerical treatment of each stage in Table \ref{tab:setup} and explain
them in detail in the following subsections. A summary table of the different sets of
star formation simulations is provided in Table~\ref{tab:sets}.

\subsubsection{Collapse phase}
\label{sec:collapsephase}

We start the simulation with an uniform sphere of gas modeled using an
isothermal equation of state to emulate the cooling of molecular
clouds in a simple and cheap way.  Such an approximation has been
widely used in star formation simulations \citep{Klessen1998,
  Klessen2000, Heitsch2001, Li2003} to avoid the inclusion of
radiative cooling recipes that are computationally expensive. Furthermore, the isothermal
regime breaks down at very high densities ($\gtrsim 1.5\times 10^{-14}\,{\rm g\,cm^{-3}}$, see \citealt{MacLow2004})
which are not being reached in the present work (see below).
We set up the initial velocity of the SPH particles by creating an artificial turbulent
velocity field in Fourier space with a energy power spectrum of $P(k) \propto
k^{-\alpha}$ with $k=\vert \vec{k} \vert$ as the three dimensional wavenumber.  To
recreate the macroscopic structure observed in star forming regions we choose a power law
of $\alpha=4$, so that energy perturbations are distributed mainly on the large scales.
We populate the $k$ spectrum with integer wavenumbers from $k=1-128$.  Then the Fourier
space velocity perturbations are transformed to 3-D real space using the inverse Fourier
transform. This results in a three-dimensional grid of $N_{\rm grid}=128^3$ cells as the
velocity field.  Then the velocities of each SPH particle are linearly interpolated from
the grid.  The velocity of the SPH particles is only set up as initial condition and no
additional energy injection is provided later, i.e.\ we do not use driven turbulence.

The resulting turbulent velocity field is a combination of two extreme
fields: the compressive forcing (curl-free) and the solenoidal 
forcing (divergence-free). On average, a random field contains 2/3 in
the solenoidal modes and 1/3 in the compressive modes \citep[see][for
details]{Federrath2009}.  Different amounts of energies in the
different modes have strong consequences in the characteristics of the
final distribution of the gas, and therefore they may affects the
final stellar distribution obtained. 

To check how much the final stellar distribution are affected by the
different modes of turbulence, we set up the initial velocity fields
in three ways: Pure compressive modes (curl-free), pure solenoidal
modes (divergence-free) and random (mixed). 

\subsubsection{Star formation phase}
\label{sec:starformationphase}

In order to avoid strong dynamical encounters -- we want to isolate effects of gas
expulsion -- only equal mass particles are formed. This is not possible to achieve with
the use of standard recipes, e.g. mass accretion by sink particles, because sinks can be
ejected from the gas-rich regions of the cluster before obtaining the desired mass.
Therefore we use an ad hoc star formation recipe forming stars instantly skipping the
accretion phase.

A gas particle $i$ and its $N_{\rm nb}-1$ nearest neighbors are combined into a single
star particle if: 1) The smoothing length $h_{\rm i} < h_{\rm crit}$
and 2) The $N_{\rm nb}$ gas particles (including $i$) are gravitationally bound.
The position and velocity of the new star corresponds to the position and velocity of the center of
mass of the combined gas particles. If two gas particles that fulfil these conditions are
on each others $N_{\rm nb}$ list, they are combined into the same star particle.
The density of gas particles is updated just after the new star forms to account for the
empty space left behind by the combined gas particles, i.e. the density in the region
decreases.  

We used $h_{\rm crit}= 0.0018\,$pc, roughly corresponding to a density threshold of
$\rho_{\rm crit}=1.34 \times 10^{-15}\,{\rm g\,cm^{-3}}$. {The Jeans
mass at that density is 0.007$\MSun$ but to properly resolve fragmentation we need
$1.5N_{\rm nb}$ per Jeans mass i.e.\ the minimum Jeans mass we are able to properly
resolve is 0.75$\MSun$ \citep[][]{Bate2002}}. This is below our resolution limit, and
fragmentation occurring in our models does therefore not depict the actual physical
process of fragmentation correctly (see Sec.  \ref{sec:icdiscussion}). Anyhow, the SPH
scheme has been shown to be stable enough to not produce artificial fragmentation even if
resolution is very low \citep{Hubber2006}. We choose $h_{\rm crit}$ to be as small as
possible without slowing down the simulations considerably.

When at least 2 stars have formed, we initiate the \textsc{Bridge}
scheme
\footnote{This is due only to a technical problem.  A code like 
\textsc{Ph4} cannot calculate forces for only 1 particle.  Before
starting the \textsc{Bridge} scheme, forces for the only present star
are evaluated by the hybrid code \textsc{Fi} in a tree scheme until 
another star is created} 
as described in Section~\ref{sec:setup}. 

This phase ends when we match the desired SFE of 20\%, i.e., when 1000
equal mass particles have formed. This allows a direct comparison with our
previous studies in which the same number of stars are simulated.

\subsubsection{Embedded phase}

The duration of the ``Collapse'' and ``Star formation'' phases is different for each
cloud. We therefore take the resulting distributions as initial conditions, and we define
the time when the star formation phase ends as $t=0$. We expel the gas in some
simulations at this time and so, for these objects, the embedded phase is skipped. For
the simulations where we choose to expel the gas later, it is not possible to continue
with the same treatment for the gas without forming more stars, since the isothermal EOS
does not provide pressure support for the cloud.  Therefore, we need to stop the
collapse. A more realistic way would be to include heating recipes in the simulations.
However, those recipes are numerically expensive. We therefore choose two extreme ways to
artificially stop the gas collapse, and the further formation of stars.  The first way is
to change the EOS from isothermal to adiabatic with an adiabatic index of $\gamma=5/3$.
In this way, since there is no cooling, thermal pressure stops the collapse. The second
way is to simply turn off the self gravity of the gas, leaving the interactions between
the stars and gas intact.  Both ways are in principle unphysical. However, it is not
clear how the gas should realistically behave during this phase since, in real star
forming regions, there are many complex physical processes involved, e.g.\ stellar
feedback, stellar winds, magnetic fields among others, which we try to avoid to include
in our simulations.  Due to these dissipative processes, it is very unlikely that the gas
forms further dense clumps inside the stellar cluster, instead it will disperse. In the
case of the adiabatic EOS, we see that the gas stays clumpy, and the largest contribution
to the potential of the cluster comes from the gas, i.e.\ this treatment mimics one
extreme.  By turning the self-gravity of the gas off, the gas disperses and would
eventually leave the cluster. However, since the velocity gained in the collapse phase is
not enough for the gas to leave the cluster in the maximum time of 2 Myr that we choose
to evolve the embedded phase, this treatment leads to the opposite extreme -- a maximum
dispersal of the gas, without leaving the region of interest.

In both cases, we expel the gas at 1 and 2 Myr after star formation has stopped.
Hereafter, we will refer to simulations with an adiabatic EOS for the gas as AEOS
simulations, and simulations with the self gravity of the gas turned off as SGO
simulations. 

\subsubsection{Gas free phase}

After gas expulsion, the gas is not present anymore and we only follow the  evolution of
the stars using the code \textsc{Ph4} alone.  We follow the evolution of the stars for 15
Myr after gas expulsion.  At this point, we measure the bound mass fraction of the
biggest clump formed in the simulation, using a method based on the iterative measure of
the mean velocity of the bound mass. We call this method the ``Snowballing Method''.
\citep[see][for a brief description, a full description of the method will be published
in Farias et al.\ in prep.]{Smith2013a} 

We perform 10 realizations for each different numerical treatment of the gas in the
embedded phase (i.e.\ either with an adiabatic equation of state or when turning the self
gravity of the gas off), 10 for each initial turbulent velocity field and 10 for each gas
expulsion time $t_{\rm exp}$ at 0, 1 and 2 Myr after the embedded phase begins.  Table
\ref{tab:sets} summarizes each of the sets for which 10 simulations were made with a
different random seed, which sums up to a total of 150 simulations.  

We run the simulations using 40 to 50 cores for the hydrodynamical integrator \textsc{Fi}
and 10 cores for the N-body module \textsc{Ph4}.  The most expensive simulations (i.e.\
the ones where the embedded phase is evolved for 2 Myr) take between 2 to 3 hours each to
complete. 

\subsubsection{On the simplicity of the star formation recipe}\label{sec:icdiscussion}

In this study we use a very rough and simplistic star formation recipe.  We emphasize
that we want to obtain an arbitrary substructured cluster on which we will test how stars
respond to gas expulsion.  Even though we would like to reproduce star formation
properly, we are limited by the computational power available to us.  A ``realistic''
sophisticated star formation recipe would exhaust it in a couple of simulations.  In this
study statistics is crucial, since much can change from one hierarchical distribution to
another, and thus we sacrifice accuracy in the star formation recipe to obtain a large
sample of simulations.  Some simplifications where made to avoid the inclusion of
additional physics, like the absence of stellar feedback, stellar evolution and the
production of only equal mass particles. The effects of this last one is being studied in
a parallel work. 

We note that using the described prescription we obtain the desired SFE in about
$\sim$1.2 $t_{\rm ff}$, where the initial free fall time is $t_{\rm ff}\approx0.6$ Myr.
This is a timescale comparable to more sophisticated star formation recipes \citep[see
e.g.][]{Bate2009}, where they obtain a SFE of $\sim$38\% in about 1.5 $t_{\rm ff}$ in
their biggest simulation.  However, this timescale may be influenced by the nature of the
initial conditions in the gas, like the properties of the initial velocity field.  A more
quantitative timescale for comparison with sophisticated star formation recipes would be
the star formation rate per free fall time SFR$_{\rm ff}$, which is defined as
\begin{eqnarray}
  {\rm SFR}_{\rm ff} &=& \dot{M_*}t_{\rm ff}/M_{\rm gas,i}
\end{eqnarray}
\citep{Krumholz2007}.
We have found that the mean SFR$_{\rm ff}$ in our simulations is $\sim 0.33$, where we
have used the initial free fall time and the mean $\dot{M_*}$ of our simulations in the
estimation.  This value is almost the same as in \citet{Price2009} for simulations
without magnetic fields and no radiative feedback.  This means that our simulations do
not have effects caused by a too fast (or slow) star formation phase.  These timescales
were achieved by using a density threshold beyond our resolution limit.  If we would
choose the threshold according to our resolution limit, then star formation would happen
too fast (since simulations reach these densities earlier), before the cloud forms the
filamentary structure observed in star forming regions.  We obtain the desired structures
at the cost of letting the simulation go beyond the recommended accuracy.  This means
that we can simulate large scale structure (in the gas and stars) confidently, but not in
the small scales (systems less massive than 0.75$\Msun$). The potential consequence is that we
do not resolve the formation of primordial binaries or multiple systems properly.
However, effects of binaries are only important when including an IMF, which effects will
be considered in a future study.  Another consequence is that gas fractions inside small
sub clusters may not be correctly modeled with uncertainties stemming from the
combination of resolution of the gas and the absence of accretion in the recipe.
However, the correct fraction of gas that subclusters should have is unclear, as is how
exactly stars and gas are coupled in the star cluster formation process.  

We emphasize once again, that we do not seek to achieve a completely correct star
formation simulation, our goal is to test the response of embedded star distributions to
instantaneous gas expulsion, when gas and stars are both in substructured distributions,
in contrast with our previous studies where this has been tested assuming a spherical
distribution for the gas.  By changing the treatment of the gas after all stars have form
(the embedded phase) we create different possible scenarios in which we can remove the
gas and measure the outcome.  

\begin{table}
  \centering
  \begin{tabular}{l || l c }
    Set & Velocity field & $t_{\rm exp}$ [Myr] \\ \hline 
    \texttt{ NEP\_c}  &  compressive    &  0 \\     
    \texttt{ NEP\_m}  &  mixed          &  0 \\
    \texttt{ NEP\_s}  &  solenoidal     &  0 \\
    \texttt{ AEOS1\_c}&  compressive    &  1 \\
    \texttt{ AEOS1\_m}&  mixed          &  1 \\
    \texttt{ AEOS1\_s}&  solenoidal     &  1 \\
    \texttt{ AEOS2\_c}&  compressive    &  2 \\
    \texttt{ AEOS2\_m}&  mixed          &  2 \\
    \texttt{ AEOS2\_s}&  solenoidal     &  2 \\
    \texttt{ SGO1\_c} &  compressive    &  1 \\
    \texttt{ SGO1\_m} &  mixed          &  1 \\
    \texttt{ SGO1\_s} &  solenoidal     &  1 \\
    \texttt{ SGO2\_c} &  compressive    &  2 \\
    \texttt{ SGO2\_m} &  mixed          &  2 \\
    \texttt{ SGO2\_s} &  solenoidal     &  2 
  \end{tabular}
  \caption{Summary table of the sets of star formation simulations
    performed in this work.  For each set we performed 10 simulations
    with different random seeds.  Column 1 the name of the set that is 
    related to the numerical treatment of the gas during the embedded
    phase, i.e.\ \texttt{NEP} stands for \emph{no embedded phase},
    \texttt{AEOS} for \emph{adiabatic equation of state} and
    \texttt{SGO} for \emph{self gravity off}.  Column 2 shows the
    nature of the initial velocity field for the gas and column 3 is
    the gas expulsion time measured after the 1000 stars form, i.e.\
    from the beginning of the embedded phase.} 
  \label{tab:sets}
\end{table}

\section{Results}
\label{sec:results}
\subsection{A smooth gas background}
\label{sec:plummer}

\begin{figure}
  \centering
  \includegraphics[width=\columnwidth]{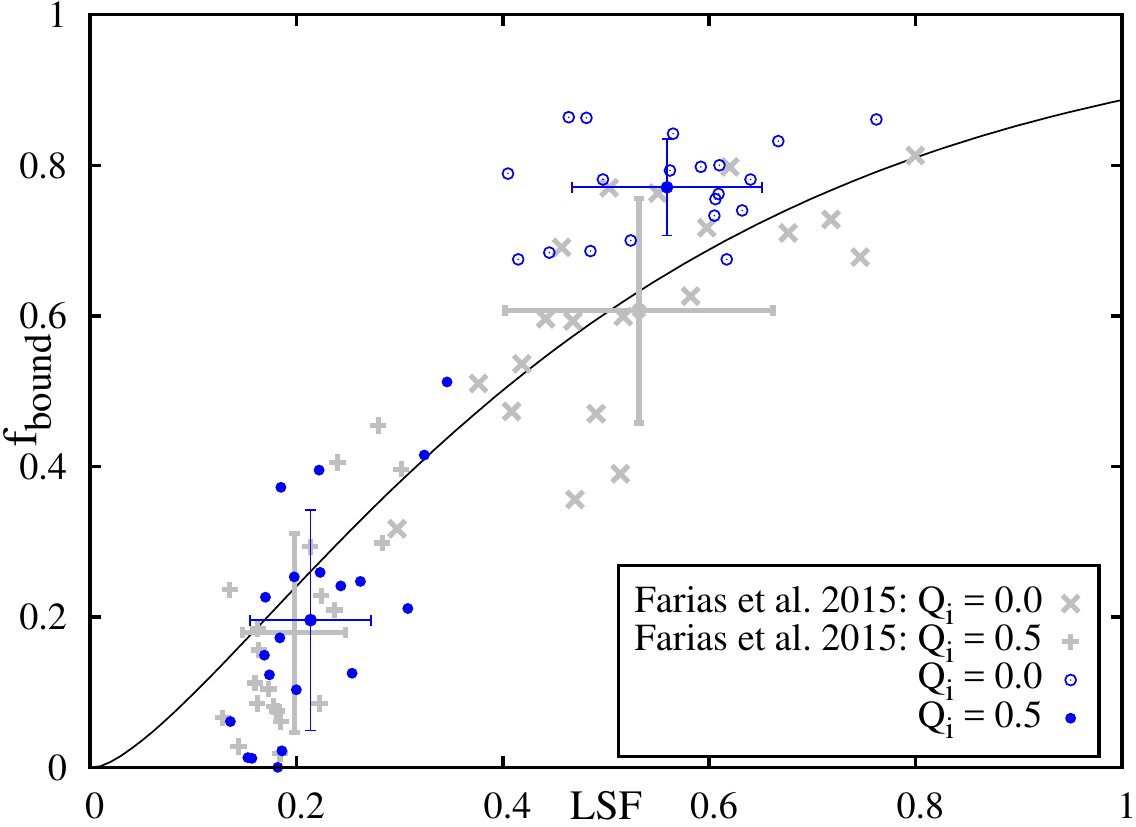}
  \caption{The $\fb$-LSF trend for fractal clusters embedded in a
    Plummer sphere of gas.  Gray symbols are simulations with a static
    background potential \citep{Farias2015} and blue symbols are
    simulations using a live gas background.  In the simulations shown
    in this plot gas is expelled at the same dynamical time, i.e.,
    when $Q=0.5$ and $Q$ is rising for the second time in the cluster
    evolution.  Black line represents the \citet{Farias2015} model,
    i.e., Eq. \ref{eq:fbound} assuming $B/A=1$ (or $\eta=1/$LSF).} 
  \label{fig:plummertrend}
\end{figure}

\begin{figure*}
  \center
  $\begin{array}{lr}
    \includegraphics[height=2.8in]{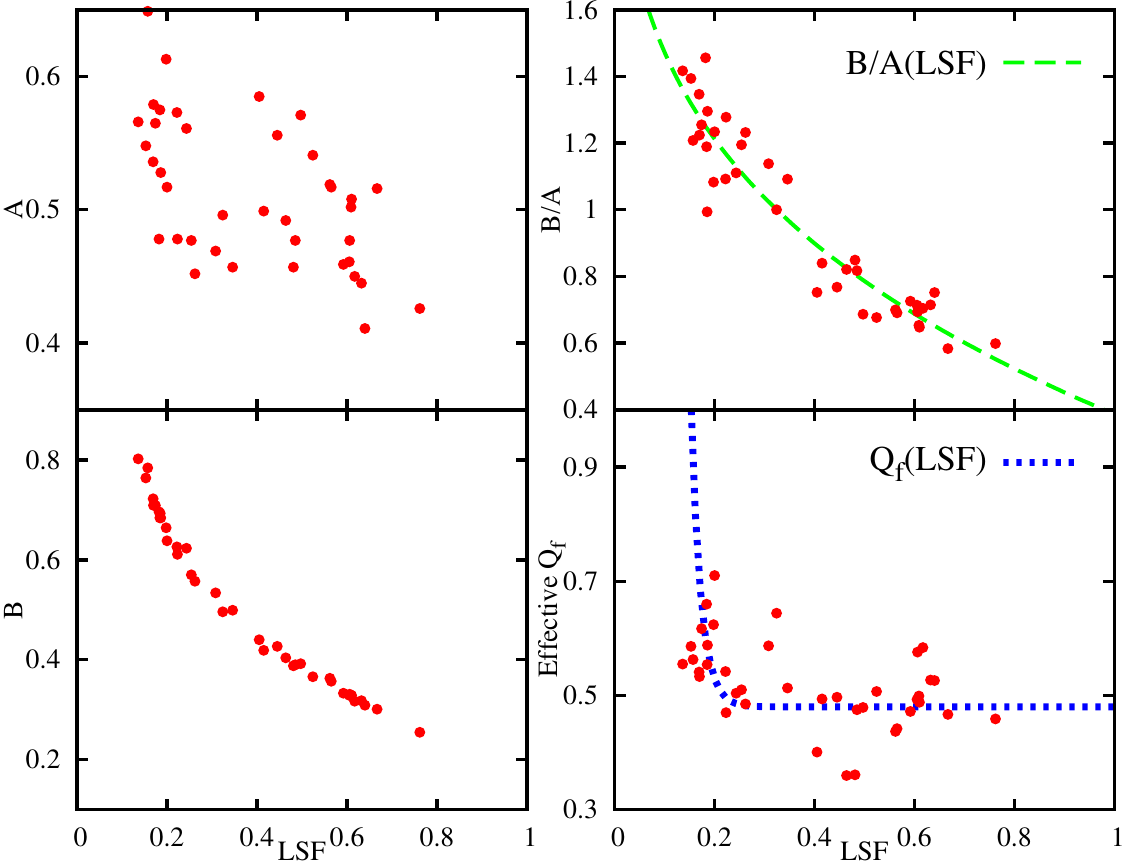} &
    \includegraphics[height=2.8in]{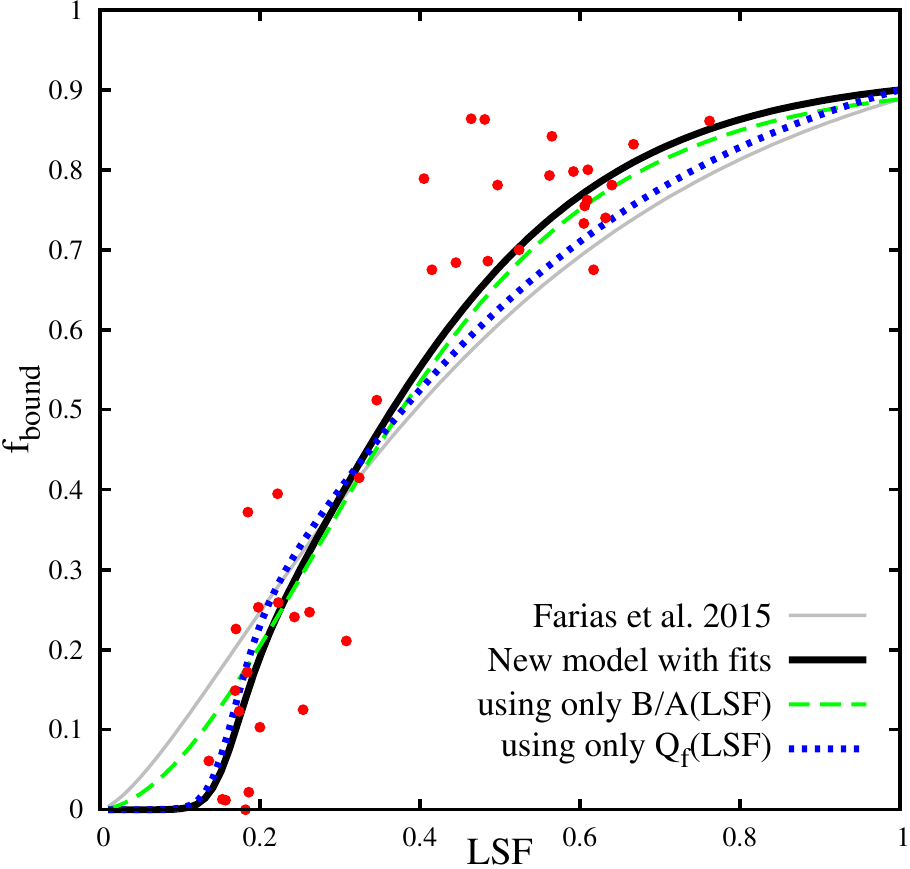}
  \end{array}$
  \caption{\textbf{Left}: The structural parameters $A$, $B$ and the
    ratio $B/A$ for simulations using a smooth live Plummer sphere as
    background. While $A$ is similar for simulations $B$ is highly
    dependent on the LSF where stars with low LSF are concentrated in
    the center and stars are not affected by the gas since most of it
    is outside the cluster.  This dependency is reflected in the $B/A$
    ratio and finally in the bound fraction  of the cluster.  We also
    show that the effective $\Qf$ is another source of error in the
    $\fb$-LSF trend.  To quantify the effect we show two fits to these
    parameters in a dashed green line for $B/A$ and a dotted blue line
    for the effective $\Qf$. \textbf{Right}: We apply the previous
    fits showing that the particular shape of the $\fb$-LSF described
    in previous papers is a consequence of a combination of the
    particular geometry of the systems evolved and also our ability to
    measure $\Qf$ correctly and is not necessarily a universal trend.} 
  \label{fig:fixplummer}
\end{figure*}

By using a smooth Plummer sphere of gas and expelling the gas when the systems have
exactly $\Qf=0.5$ we obtain the same trend as in \cite{Farias2015} (see
Fig.~\ref{fig:plummertrend}).  The main difference in both cases is a slight offset in
the region of LSF that clusters populate.  This is because if interactions between stars
and gas are possible, stars loose energy in the interaction and sink to the center,
raising the LSF.  However, the change is not significant and the $\fb$-LSF trend is the
same as in \citet{Farias2015}. 

Even though the trend is the same, it is quite obvious that the simple model of
\citet{Farias2015} overestimates $\fb$ at low values of LSF ($\sim 0.2$) and the sample
of live gas simulations appears to survive better than the model expects for LSF $>$ 0.4.
We found two main reasons that give the $\fb$-LSF relation its particular shape.  The
first one is related to the basic assumptions in the model of \citet{Farias2015}. To
simplify the maths, it was assumed that the stars and gas are distributed in a similar
way ($B/A\approx1$).  We make use of the structural parameters $A$ and $B$ to show how
far simulations are from this assumption and also how much this affects the estimations.
We note that we are not suggesting to measure such values in observable star clusters,
and this is just an illustrative experiment.  While the $A$ parameter is relatively
similar for all simulations (top left panel in Fig.\ref{fig:fixplummer}), the $B$
parameter is highly dependent of how the stars are distributed inside the background gas
(bottom left panel in Fig.\ref{fig:fixplummer}).  We show the structural parameters as a
function of the LSF.  $A$ is similar for all the simulations since at that point the
level of substructure and shape of the clusters are similar, however $B$ shows a strong
dependency of the LSF.  The reason is that at high LSFs stars are concentrated in the
center and most of the gas is in the outer layers of the cluster (then $B$ is low).  On
the other hand, at low LSF  the cluster is expanded and there is more gas inside the
cluster ($B$ raises).  Therefore, the ratio $B/A$ is highly dependent on the  LSF (top
middle panel), and thus Eqs. \ref{eq:eta} and \ref{eq:fbound} imply that clusters survive
better when $B/A < 1$ and the opposite when  $B/A > 1$.  We show a fit to this ratio
(green dashed line in Fig.~\ref{fig:fixplummer}) and we show how this effect affects the
model of \cite{Farias2015} in the right panel of Fig.~\ref{fig:fixplummer} as the green
dashed line.  We see that this variation clearly explains the shape of the $\fb$-LSF
trend at high LSF and also at low LSF. However, the effect of the LSF-dependent
$B/A$-ratio is not strong enough to explain why star clusters don't survive with LSFs
below 0.2, as predicted by the \cite{Farias2015} analytical model.

\begin{table*}
\centering
\begin{tabular}{l c c c c c c c }\toprule
        & $R_{\rm h}$[pc]  & LSF   & $R_{\rm max}$[pc]& SFE$_{R_{\rm max}}$ & $\cal{C}$ & $\Qi$ & $\sigma_*$[km/s]\\\hline
\multicolumn{2}{l}{Plummer Background}    \\ 
Cold               & $0.9\pm0.1 $ &$0.19\pm0.03$& $1.5 $          &   $0.2 $              & $0.39\pm0.06 $& $0.0 $  &  $ 0       $      \\
Warm               & $0.9\pm0.1 $ &$0.19\pm0.03$& $1.5 $          &   $0.2 $              & $0.39\pm0.06 $& $0.5 $  &  $1.0\pm0.1$      \\ \hline
\multicolumn{2}{l}{Turbulent Setup}    \\ 
Divergence Free    &$0.5  \pm0.3 $  &$0.6 \pm0.2 $& $2.5  \pm0.9 $  &$0.24 \pm0.03$      &$0.28\pm0.07$ &$0.31 \pm0.06$ &$1.1  \pm0.2 $ \\
Mixed              &$0.4  \pm0.3 $  &$0.6 \pm0.2 $& $2.7  \pm0.9 $  &$0.23 \pm0.02$      &$0.27\pm0.08$ &$0.31 \pm0.07$ &$1.1  \pm0.2 $ \\
Curl Free          &$0.6  \pm0.4 $  &$0.5 \pm0.2 $& $2.5  \pm0.9 $  &$0.23 \pm0.03$
&$0.24\pm0.08$ &$0.27 \pm0.06$ &$0.9  \pm0.2 $ \\\bottomrule
\end{tabular}
\caption{A comparison between the initial conditions generated by the
  turbulent setup and simulations with fractal distributions embedded
  in a Plummer background.  Values are means with respective standard
  deviations for each set of simulations described in column 1.
  Column 2 shows the initial half mass radius of the stellar
  distribution in parsecs, column 3 the initial local stellar
  fraction, column 4 the radius containing all the stars in
  parsecs, column 5 the star formation efficiency measured at $R_{\rm
    max}$, column 6 the amount of primordial substructure measured by
  the $\cal C$ parameter, column 7 the initial virial ratio and column
  8 the velocity dispersion of the stellar component.} 
\label{tab:ic}
\end{table*}

The second reason of the particular shape of the trend is our ability to measure $\Qf$.
The virial ratio is highly dependent on the frame of reference.  While the potential
energy is not, the kinetic energy in the cluster is highly dependent of what we choose as
the mean velocity of stars in the cluster.  The simplest way is to use the  mean velocity
of the whole star distribution, and this generally works fine if we are in the ballpark
of high LSF values.  However, a more correct characteristic velocity for the cluster
would be the mean velocity of only the stars that will actually remain bound.  Of course
we cannot know this last velocity since we would need to know exactly what particles will
remain bound a priori.  But we know that in general, when a high fraction (or at least
representative) of the stars remains bound after gas expulsion, the mean velocity of the
whole distribution is close enough to the velocity of the bound cluster, and the
calculated $\Qf$ is then representative.  However, when $\fb$ is low, there is a lower
chance that both velocities coincide.  We call the virial ratio measured with the
velocity of the stars that finally will remain bound the \emph{effective} virial ratio.
In reality, it is not possible to measure such value, but in our simulations we have all
the information that we need to track the bound particles back and measure their mean
velocity.  The bottom middle panel on Fig. \ref{fig:fixplummer} shows the effective $\Qf$
as a function of the LSF.  At low LSFs, the velocity of the system is not representative
of the one of the bound stars.  While globally the star distributions have $\Qf=0.5$ by
design (see our criteria for the gas expulsion time), effectively the bound system has
$\Qf>0.5$ resulting in an over prediction of $\fb$. In order to illustrate how much the
model of \citep{Farias2015} is affected by these effects, we calculate the relation
between LSF and $\fb$ with the fits shown in the middle panels of Figure
\ref{fig:fixplummer} as inputs.

Both effects are important in different regimes, and the particular shape of the
$\fb$-LSF trend is a combination of both. But more importantly, this simple fitting shows
that the shape of the $\fb-LSF$ trend is not general and depends on how the stars and gas
are distributed with respect to each other. Predicting the bound fractions seems to be
quite a difficult task, especially if gas and stars remain substructured and have not had
time to become a more spherical distributions, and also if the bound entity is small.

We will return to the topic of predicting the survivability of star clusters to
instantaneous gas expulsion further down in the text.  Here we want to stress that our
simulations are designed to fully explore the parameter space in LSF and $\Qf$ to be able
to compare the analytical predictions to the complete trend obtained by our simulations.
At this point, we are not concerned if the full parameter space extends beyond that
inhabited by real star clusters, but we raise this issue again in the Discussion section.

\subsection{Highly substructured gas distributions}\label{sec:substructure}

As we describe in Section~\ref{sec:starformsetup}, we expel the gas of new born star
clusters at three different times: Just after stars form (0 Myr) or after 1 or 2~Myr of
embedded evolution.  We follow the embedded evolution utilizing two very different
treatments for the gas in order to avoid further collapse.  Both approaches are likely
unrealistic. However, they represent two extremes in the possible spatial distributions
of the gas, that could have great relevance for the evolution of the stars after gas
expulsion, since the gravitational potential fields that they produce are extremely
different. We will avoid the discussion of which scenario is closer to reality for now.
We emphasize the objective of the simulations presented in this work are rather
illustrative, to show the effects of large variations in the background substructure,
rather than to attempt to match the background substructure found in real star clusters. 

\subsubsection{The new initial conditions}
\label{sec:newic}

\begin{figure*}
  \centering
  \includegraphics[width=0.98\textwidth]{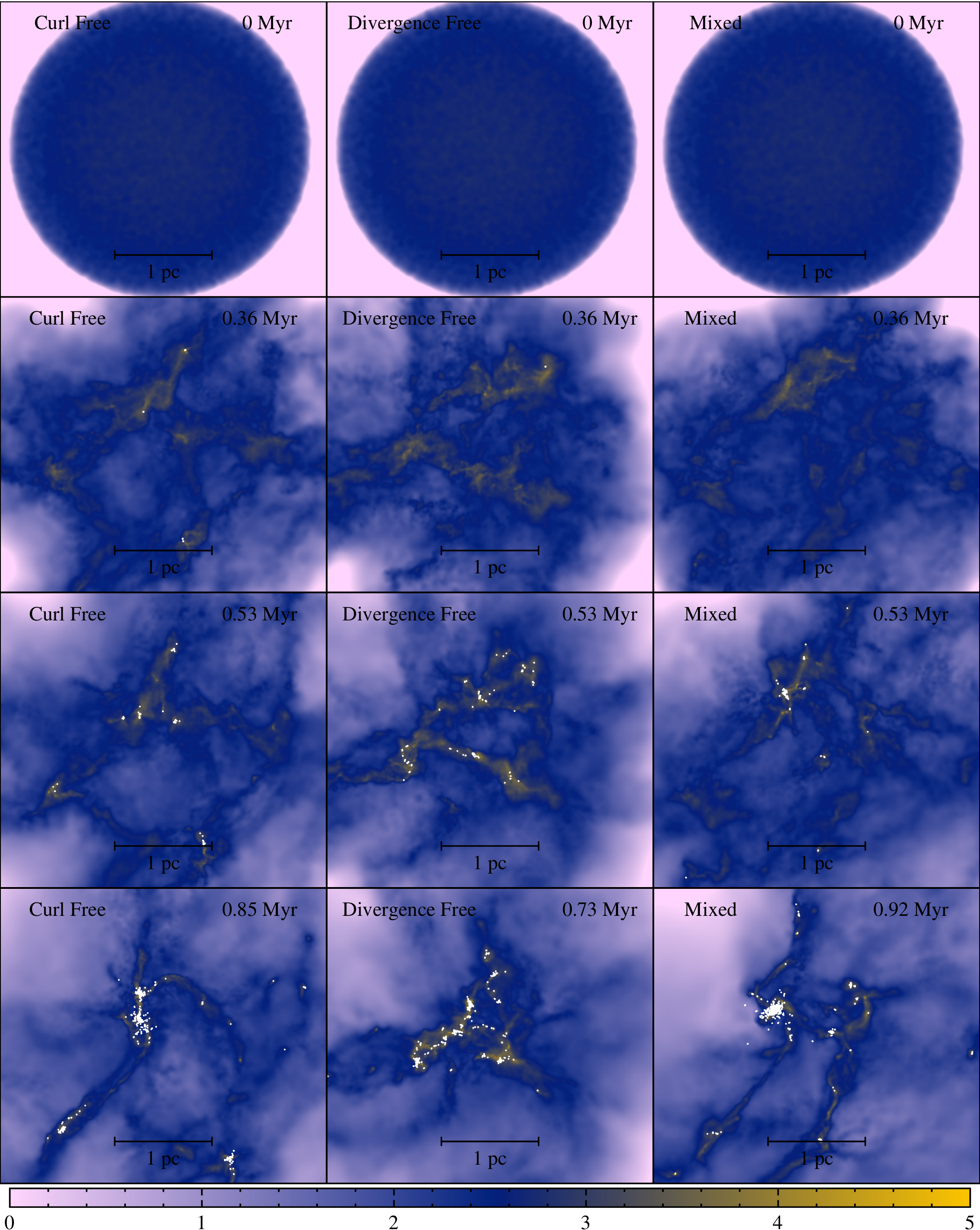}
  \caption{Evolution of the initially homogeneous turbulent molecular
    clouds until $N=1000$ equal mass stars are formed.  Simulations
    with curl-free (left column), divergence-free (middle column) and
    mixed (right column) turbulent fields are shown at (from top to
    bottom panels) 0 Myr, 0.36 Myr, 0.56 Myr and when 1000 stars are
    formed using the same random seed.  Each panel has $3\times3$ 
    pc$^2$ and the color bar represent the logarithmic column density
    measured in $\Msun$/pc$^2$. This figure, as well as the others
    column density figures in this work, have been prepared with the
    \textsc{Splash} tool developed by \citet{Price2011}.} 
  \label{fig:fractals}
\end{figure*}

\begin{figure*}
  \centering
  \includegraphics[width=0.96\textwidth]{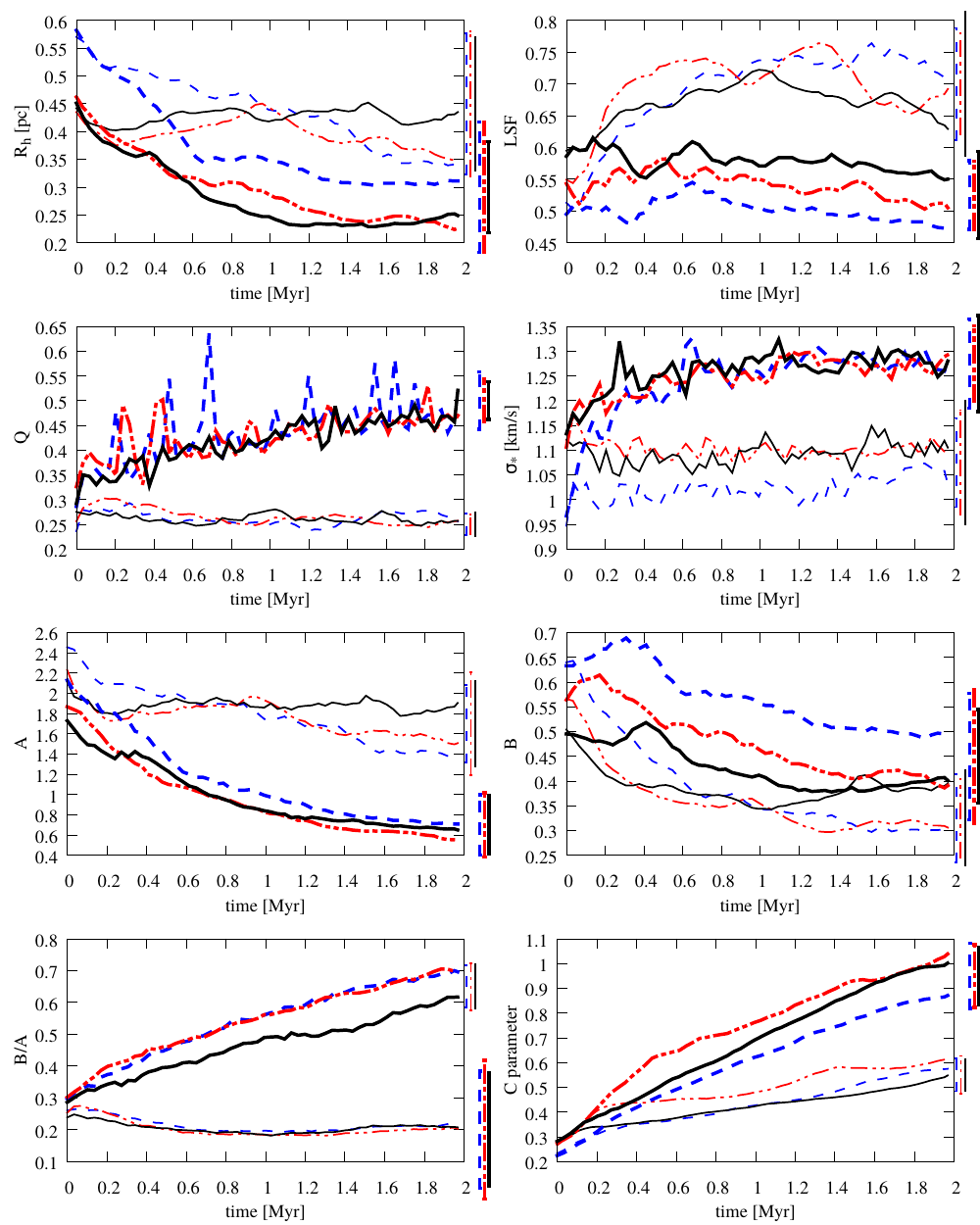}
  \caption{The evolution of the embedded star clusters for 2 Myr after
    the $N=1000$ stars form.  Values are means for each parameter of
    the 10 realizations for each setup where standard deviations been
    omitted for clarity.  From top to bottom and left to right: The
    stellar half mass radius $R_{\rm h}$, the Local Stellar Fraction,
    the virial ratio $Q$, the stellar velocity dispersion $\sigma_*$,
    the structure parameters $A$ and $B$, the $B/A$ ratio and the $\C$
    parameter. Thick lines are AEOS simulations and thin lines are
    SGO simulations. In both cases divergence free setup is shown in a
    dot dashed red line, curl-free in a dashed blue line and mixed setup in
    solid black lines. The standard deviation on each ensemble is averaged over
    time and shown at the right of each panel to represent a typical error.  We advise to
    the reader to pay attention to the minimum and maximum values of the y axes on the
    different panels.} 
  \label{fig:embedded}
\end{figure*}

The nature of the initial velocity field has a strong consequence in the substructure
formed by the gas.  While compressive motions tend to form large voids and filaments,
solenoidal velocity fields tend to form a more uniform substructure \citep[see
e.g.][]{Federrath2009}.  Hence, we split our simulations in three groups depending of the
initial velocity field: the compressive (curl free), solenoidal (divergence free) and
mixed velocity fields.  Fig.~\ref{fig:fractals} shows snapshots at different times of the
three kinds of simulations until the end of the star formation phase.  

We attempt to form systems similar to the ones in our first experiment with a Plummer
Sphere of gas.  Table \ref{tab:ic} shows a summary of some important parameters that we
compare with the initial conditions of simulations using a smooth background gas. 

To measure the level of substructure we make use of the $\cal C$ parameter\footnote{The
parameter is called $\cal Q$ parameter by the authors, however we call it $\cal C$ to
avoid confusion with the virial ratio Q.} introduced by \citet{Cartwright2004} which is
the ratio between the area normalized mean length of a minimum spanning tree joining all
the particles ($\bar{m}$) and the area normalized mean separation between particles.
Values of ${\cal C} < 0.8 $ are obtained in fractal-like stellar distribution (where a
lower $\cal C$ is obtained with smaller fractal dimensions, i.e.\ high level of
substructure) and ${\cal C}>0.8$ is obtained in spherical distributions where a higher
value of $\cal C$ matches a steeper density profile.  

Even though there is a big difference between the substructure of the gas generated by
either a curl-free or divergence-free velocity field, this is not expressed in the
resulting stellar distributions where the resulting level of substructure is very similar
in all cases.  This is because stars and gas quickly decouple, and stars tend to form
their own independent distribution through mergers of sub-groups.  This similarity is in
agreement with \citet{Lomax2015} and \citet{Girichidis2012}, where the same turbulent
modes where tested.  We obtain the same slight difference in the mean values of $\cal C$
as \citet{Girichidis2012}, with $\mean{{\cal C}_{\rm comp}} \lesssim \mean{{\cal C}_{\rm
mix}} \lesssim \mean{{\cal C}_{\rm sol}}$.  However, the differences are very small and
well within the standard errors.  

We obtain even more substructured star clusters than the $D_{\rm f} =1.6$ fractals used
in our previous studies, with ${\cal C} \sim 0.26$, comparable to fractal distributions
with $D_{\rm f} < 1.5$. 

In comparison with the initial conditions used in our previous studies, we now form
clusters with 2.5 pc radius and SFE $\sim$ 24\%.  This is slightly higher than the setup
SFE of 20\%, since there is always some gas outside the maximum radius of the cluster
$R_{\rm max}$.  Distributions using the turbulent setup form with a higher LSF. This is a
consequence of the shape of the gas, which is distributed in filaments around the cluster
rather than concentrated in the center of the stellar distribution.  Furthermore, we
obtain smaller half mass radii meaning that in general our stellar distributions are more
centrally concentrated in comparison with the fractal method described by
\citet{Cartwright2004}. 

\subsubsection{Embedded evolution}
\label{sec:embedded}

The embedded evolution strongly depends on the numerical treatment of the gas, or more
accurately, on the behaviour of the gas.  It is important to know how the star clusters
behave under different background gas conditions since this determines their matter
distribution and dynamical state at the time of gas expulsion.

Fig.~\ref{fig:embedded} shows time evolution of the mean stellar values for the different
sets that expel the gas at 2 Myr of embedded evolution.  Note that the initial conditions
for the AEOS and the SGO simulations are the same for each turbulent setup, and
differences only depend on the background gas. 

In AEOS simulations (thick lines) gas quickly forms clumps which, as a difference with an
isothermal EOS, are thermal pressure supported.  Anywhere a overdensity exists at the end
of the star formation phase, changing the EOS to adiabatic causes the gas to quickly form
roughly spherical gas clumps in internal equilibrium that later merge into larger clumps.
The stars follow the potential generated by these clumps causing the star cluster half
mass radius to decrease (first row, left panel), however the LSF remains roughly constant
and does not rise like in the static background case (first row, right panel), since gas
is also being concentrated in the center.  Interactions caused by the mergers help the
star cluster to reach equilibrium, as we can see in the $Q$ panel of
Fig.~\ref{fig:embedded} (second row, left panel) the virial ratio increases roughly
linearly and usually 2 Myr are enough for these star clusters to reach equilibrium.  In
contrast, in SGO simulations (thin lines) gas disperses instead of forming clumps,
overdensities are not so strong and stars do not have a clearly defined potential where
to merge. Therefore, the local free fall time, and also the crossing time of the region
(since stars have small velocities; see second row, right panel) is longer and thus the
timescale that the cluster needs to virialize is longer.  As consequence, stars do not
have time to virialize in the 2 Myr that we evolve the embedded phase. 

Despite those differences, the strength of the potential field generated by the gas is
quite similar (see parameter $B$ in Fig.~\ref{fig:embedded}; third row, right panel),
with the AEOS simulations being stronger, however the big difference we can appreciate in
the $B/A$ ratio (fourth row, left panel) comes from how the stars rearrange in both
scenarios.  The parameter $A$ (third row, left panel) summarizes the strength of the
potential generated by the stars and decreases with time for the AEOS simulations, this
is a consequence of erasing the substructures.  At 0 Myr, stars are distributed in a
fractal-like substructure and stars are generally very close to each other in comparison
with the volume of the sphere that contains them.  This raises the potential energy of
the stars, however, in AEOS simulations substructure is erased very quickly due to
mergers (as we can see in the $\cal C$ parameter panel of Fig.~\ref{fig:embedded}; fourth
row right panel).  When substructure is erased the stellar distribution spreads over the
volume decreasing the value of $A$.  Even though we can see that substructure of the SGO
simulations also decreases ($\cal C$ raises), the $\cal C$ parameter never reach the
${\cal C} =0.8$ limit that split spherical distributions from substructured
distributions, and $A$ decreases very slowly. 

As we see in section \ref{sec:model1}, star cluster survivability depends on $Q$ and the
$B/A$ ratio, which never reach values above 1 in these simulations, meaning that they can
survive better.  Considering also the low values of $Q$ for the SGO simulations, the
simulated star clusters are likely to be able to easily survive gas expulsion. We
will analyse the outcomes of gas expulsion in the next section.

\subsubsection{Survival to gas expulsion}
\label{sec:ge}

\begin{figure*}
  \center
  \includegraphics[width=0.75\textwidth]{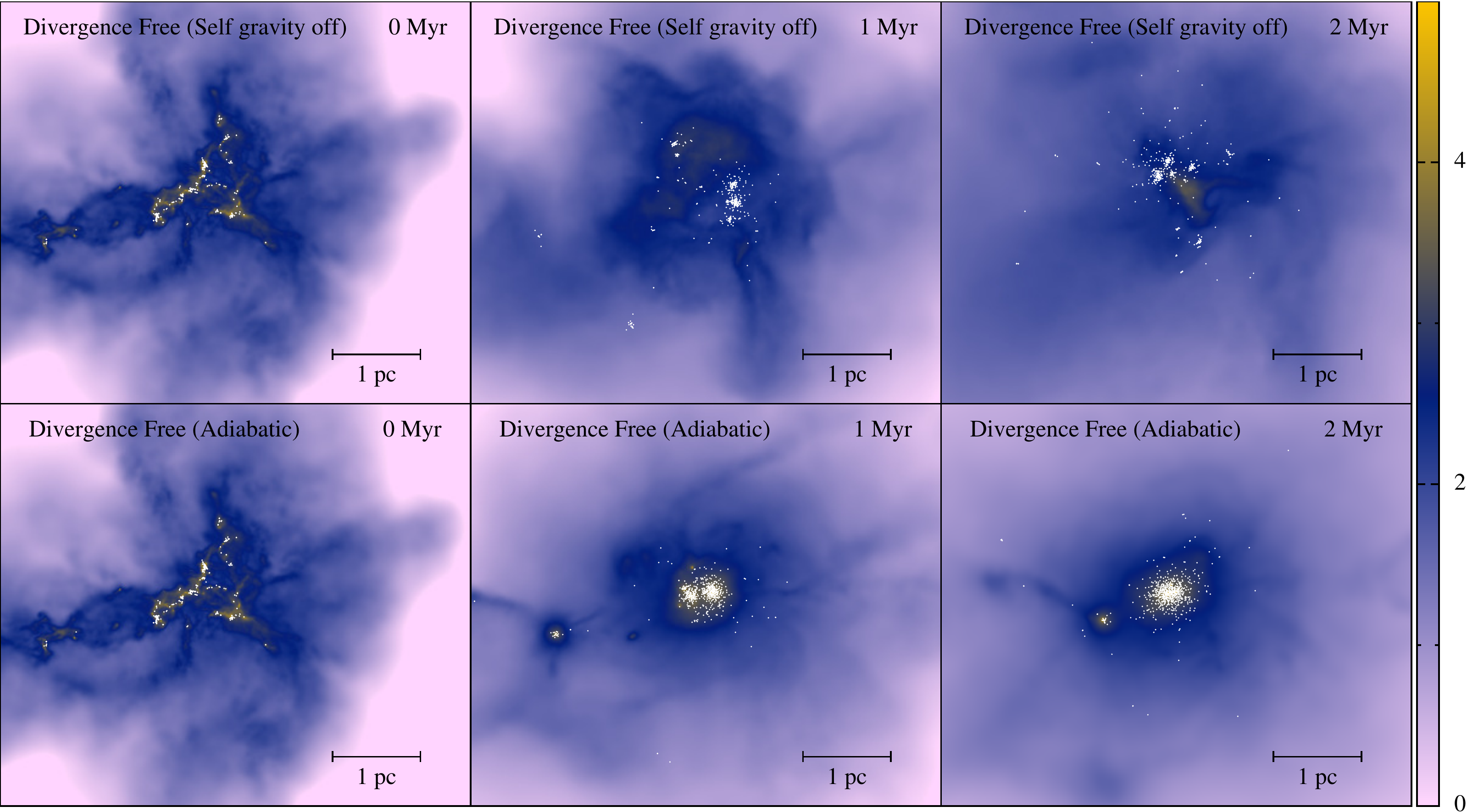}
  \includegraphics[width=0.75\textwidth]{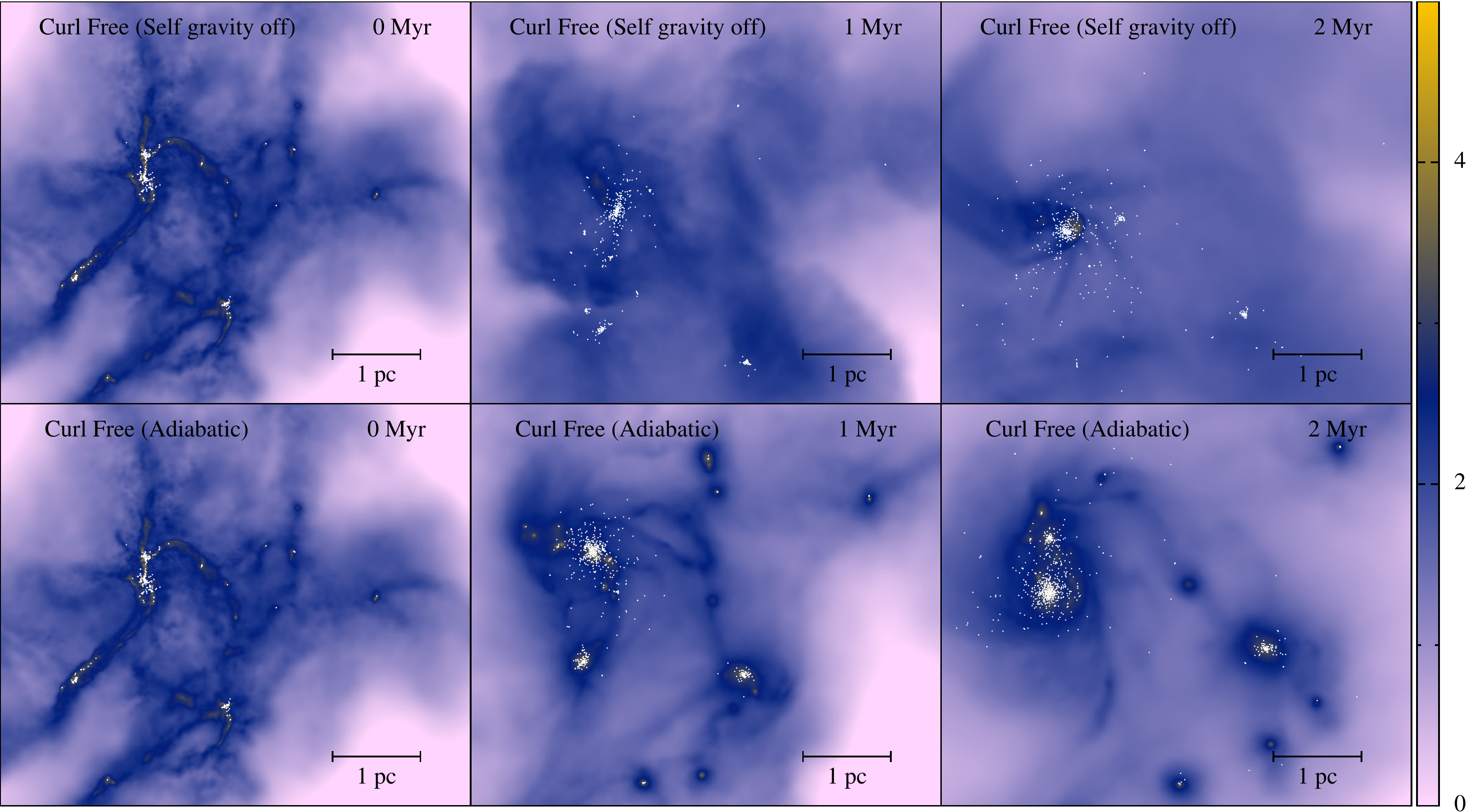}
  \includegraphics[width=0.75\textwidth]{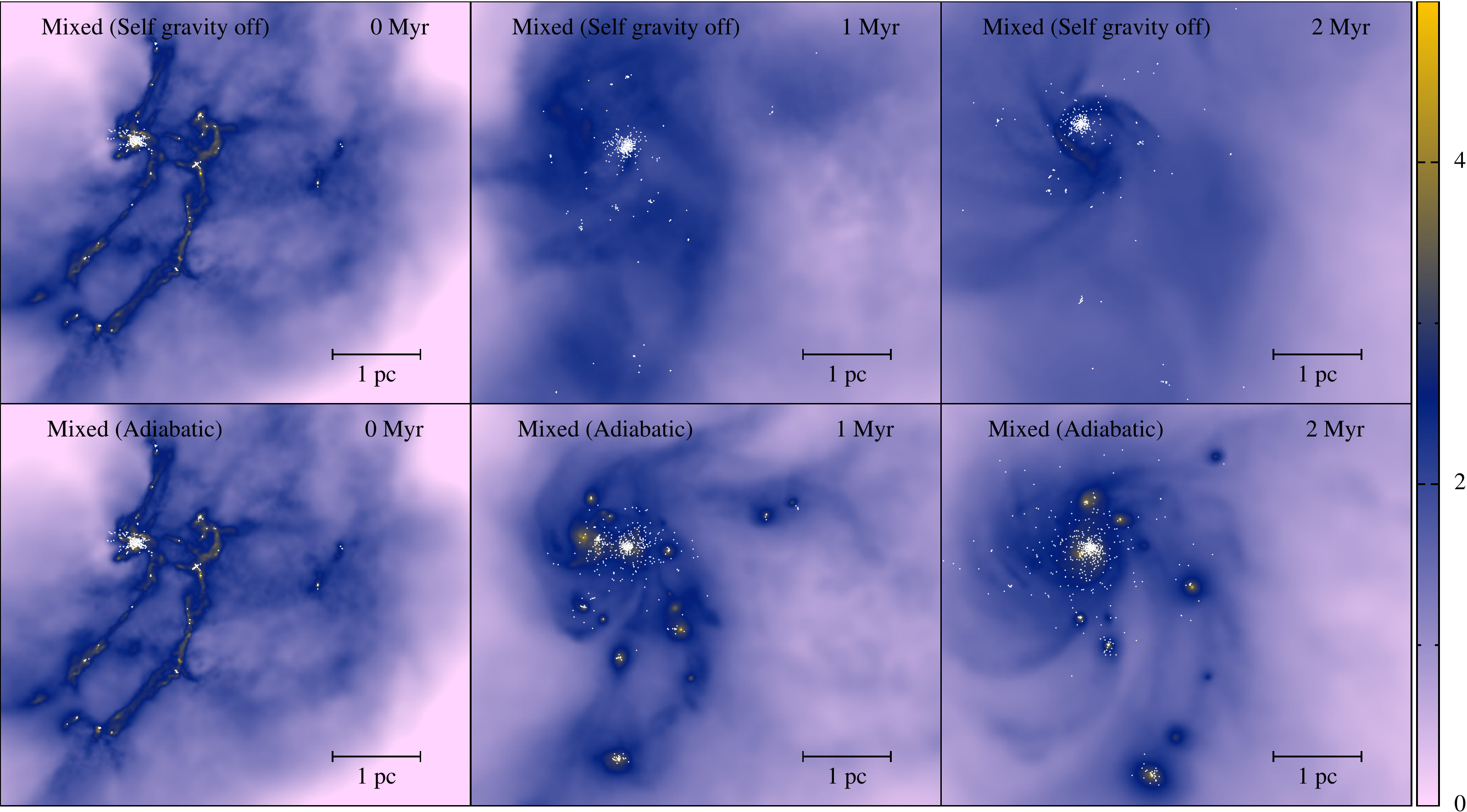} 
  \caption{Example snapshots of embedded star clusters with each
    initial turbulent velocity field at the times when gas is
    expelled.  Top panels of each 6 panel sets, shows the evolution of
    the cluster  when self-gravity of the gas is turned off (SGO
    simulations), bottom panels show the same distribution but
    evolving under an adiabatic EOS for the gas (AEOS simulations).
    Colors represent the Logarithm of the column density measured in
    $\Msun$/pc$^2$. } 
  \label{fig:snapshots}
\end{figure*}

\begin{figure*}
  \centering
  \includegraphics[width=0.92\textwidth]{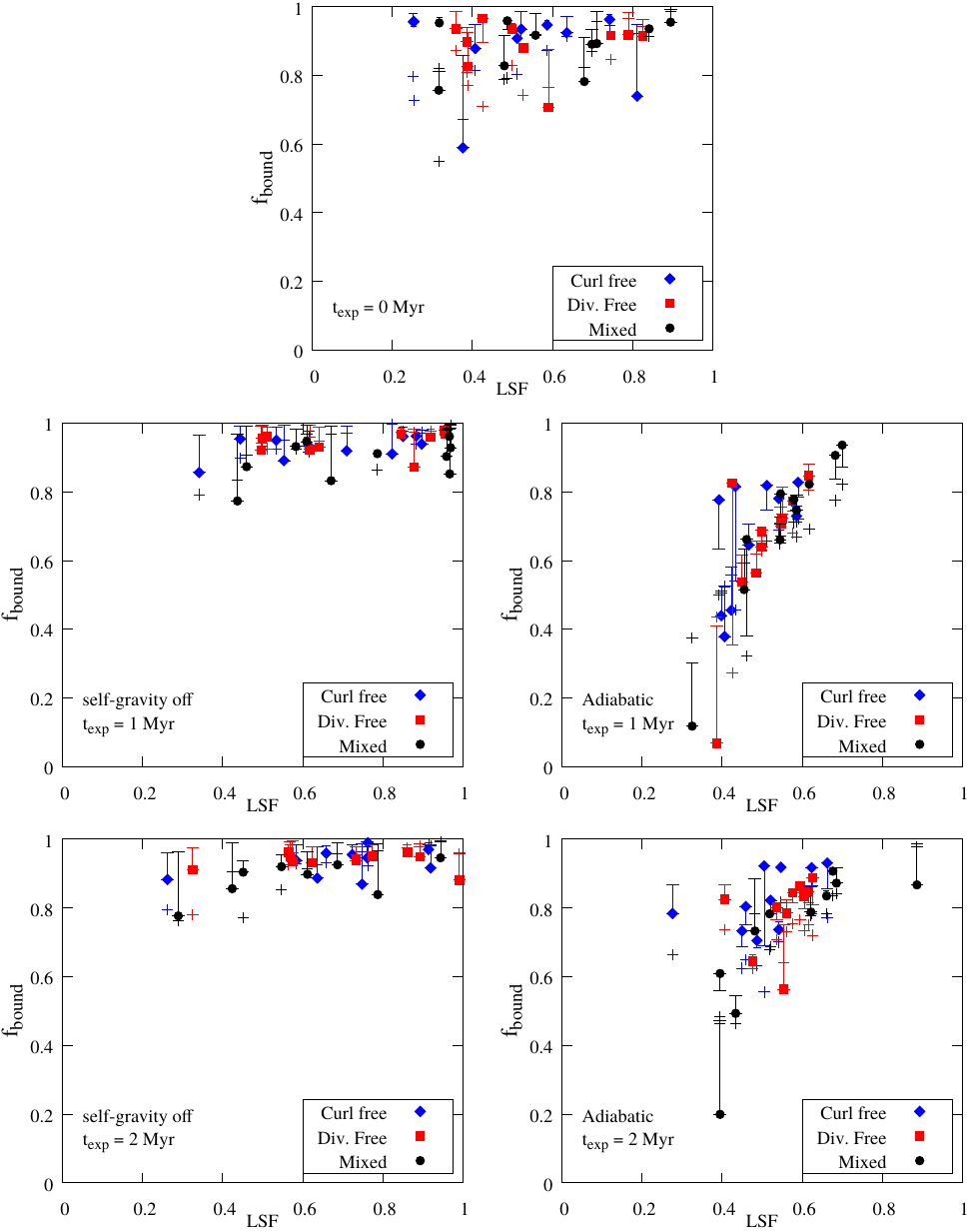}
  \caption{Bound fractions for embedded star clusters at 15 Myr from
    the moment of gas expulsion.  Different panels show sets of
    simulations expelling  the gas at 0 Myr (top panel), 1 Myr (middle
    panels) and 2 Myr (bottom) after the end of the star formation
    phase. Panels at the left are the results from SGO simulations and
    right panels show the resulting bound fractions from AEOS
    simulations.  The different natal velocity fields are shown as
    different symbols with curl-free velocity fields in blue diamonds,
    divergence free fields in red squared and the mixture of both in
    black circles.  Error bars show the prediction of the model
    presented in this work taking into account the substructure of the
    gas and the stars through the parameters $A$ and $B$.  Crosses are
    the predictions of the model without considering substructure as
    described in \citep{Farias2015}, i.e., assuming $B/A=1$.  All model
    predictions are calculated using the effective $\Qf$ discussed in
    Sec.~\ref{sec:plummer}.} 
  \label{fig:fbounds}
\end{figure*}

We expel the gas in the cluster at three different times after the embedded phase starts,
at the beginning (0 Myr), at 1 and at 2 Myr.  Then we measure the bound fraction after 15
Myr of gas free evolution.  Fig.~\ref{fig:snapshots} shows snapshots of one of our
simulations for each of the initial turbulent fields and numerical treatment of the
background gas at the moment of gas expulsion, so that we can clearly see the level of
remaining substructure in the gas and stars at each stage.  Fig.~\ref{fig:fbounds} shows
the results of the bound fraction measurement after 15 Myr of gas expulsion for each set.
We include the predictions of the analytical model described in section \ref{sec:model1},
which accounts for the independent substructure of the stars and gas as error bars (i.e.\
Eq.~\ref{eq:fbound}). We also show the prediction of the analytical model introduced in
\citet{Farias2015}, which assumes a identical distribution of mass for the stars and the
gas (i.e.\ Eq.~\ref{eq:fbound} assuming $B/A=1$). All predictions shown in
Fig.~\ref{fig:fbounds} are deduced using the effective $Q_{\rm f}$ discussed in section
\ref{sec:plummer}, however, this choice is not very important as we will see later in
this section. 

The high survival rates of simulations expelling the gas at 0 Myr and the SGO simulations
are mainly explained by the low virial ratios that the stars have and their inability to
reach virial equilibrium. The SGO simulations also have very low $B/A$ ratios, so in
general they survive gas expulsion remarkably well. The AEOS simulations show a trend
very similar to the Plummer background case presented in Section~\ref{sec:plummer}.  The
gas quickly rearranges into a spherical distribution and the stars follow the potential
well generated by the gas, forming (in general) a similar configuration than the Plummer
background case.  In this last scenario stars also have the chance to reach virial
equilibrium velocities and in general they are quite virialized in comparison to the SGO
simulations by the time we expel the gas.

The disagreement between the model introduced in Section~(\ref{sec:model1}) and the
numerical simulations is represented by the error bars in Figure~\ref{fig:fbounds}. This
reveals that the model introduced in Section~\ref{sec:model1} does a good job at
predicting the bound fraction (i.e. the error bars are small). It is, however, remarkable
that its performance predicting bound fractions is not always better than the simple
$B/A=1$ model. Thus, everything seems to be fairly well explained when measuring the
effective $Q_{\rm f}$ and the consideration of substructure does not improve the
predictions significantly.  And, as we will see later, even if we do not use the
effective $\Qf$, predictions are still good without considering the substructure.

\begin{figure}
  \centering
  \includegraphics[width=\columnwidth]{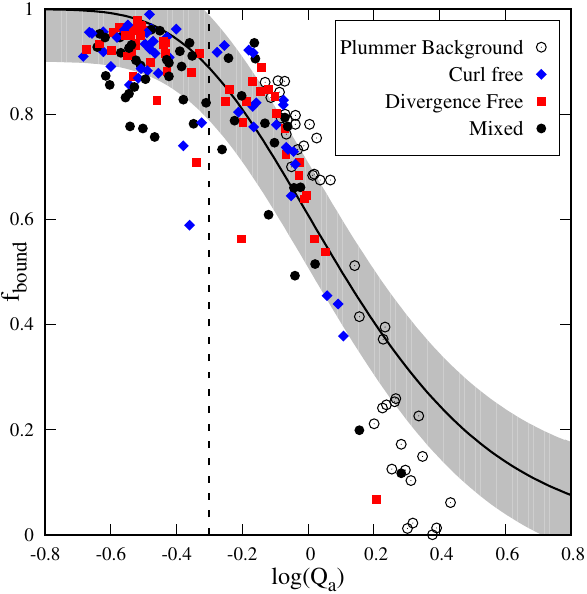}
  \caption{The post-gas expulsion virial ratio $Q_{\rm a}$ as an
    estimator of $\fb$ for all the simulations performed in this
    study.  Black solid line is the prediction of Eq. \ref{eq:fbound2}
    which is not dependent on the geometry of the system, with gray
    areas as the standard deviation from the curve.  Values of $Q_{\rm
      a}$ are effective values.  Vertical dashed line represents
    $Q_{\rm a}=0.5$ for reader's reference.} 
  \label{fig:qa}
\end{figure}

We also test the alternative approach described in Section~\ref{sec:model2}.  Since
Eq.~\ref{eq:fbound2} only depends on one parameter, the immediately post gas expulsion
virial ratio $Q_{\rm a}$ we put the results of all the simulations performed in this work
into Fig.~\ref{fig:qa}.  $Q_{\rm a}$ works quite well when estimating $\fb$ and it has
the advantage that it does not depend of the background gas, all we need is accurate
information of the positions and velocities of the stars alone (assuming instantaneous
gas expulsion).

\begin{figure*}
  \centering
  \includegraphics[width=\textwidth]{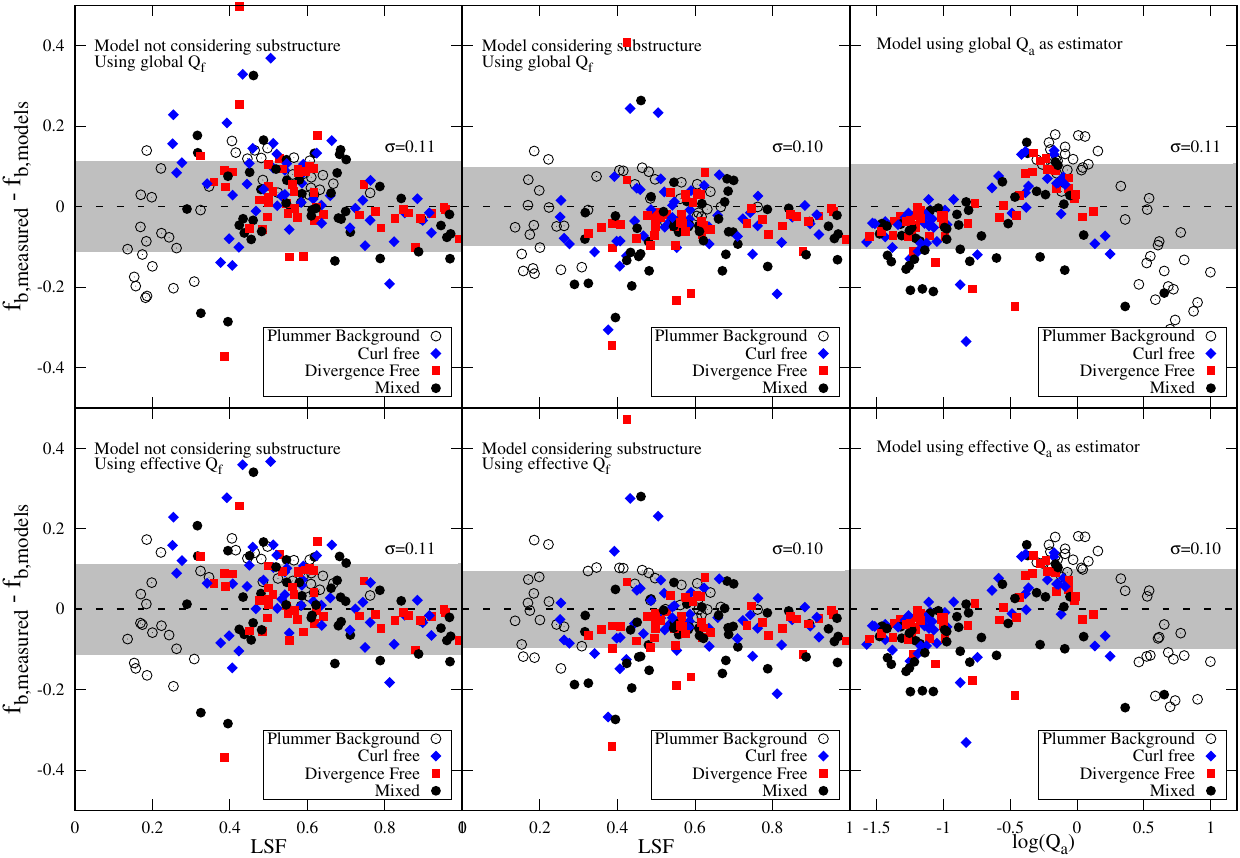}
  \caption{The differences between the analytic models and the
    measured bound fractions for all the simulations performed in this
    paper.  Top panels show the performance of the models when
    measuring $\Qf$ with respect to the global mean velocity while
    bottom panels show the performance of the models when measuring
    the effective $\Qf$ i.e.\ when using the mean velocity of the
    clump that remains bound after gas expulsion (see section
    \ref{sec:plummer}).  Left panels show the behaviour of the models
    when we do not take into account structure parameters (assuming
    $B/A=1$).  Middle panels shows results when substructure is
    included through the measure of the parameters $A$ and $B$ and
    right panels shows the performance of the model when using $Q_{\rm
      a}$ as estimator of $\fb$ (see Eq. \ref{eq:fbound2}).  Different
    natal velocity fields are shown in the same way than in
    Figs. \ref{fig:fbounds} and \ref{fig:qa}.  Simulations with a
    Plummer background gas (see Fig.~\ref{fig:fixplummer}) are shown
    as black open circles.  We measure the performance of the
    analytical models through the standard deviation from the models
    represented by the gray shaded areas.  We can see that the
    consideration of the effective $\Qf$ does not change the
    performance (due to the generally big $\fb$ obtained in this work)
    and the consideration of substructure only improves the accuracy
    of the models on a 1 percentage point level.  The use of $Q_{\rm
      a}$ does not improve the estimations either.  We show that we
    can predict the outcome of all the simulations in this work with
    10 percentage points uncertainly even in cases of high levels of
    substructure in the gas and stars.} 
  \label{fig:res}
\end{figure*}

In order to quantify the performance of different analytical models, we measure the
difference of the measured $\fb$ with the estimated $\fb$ and we calculate the standard
error for all the simulations performed in this paper.  We consider six flavours of the
analytical models introduced in Section~\ref{sec:results} and in \cite{Farias2015}. These
six flavours come about by considering three sets, namely one set where substructure is
ignored (i.e. $B/A=1$), one set where the substructure is measured through the parameters
$A$ and $B$, and one set where the clusters are characterized by their $Q_{\rm a}$. For
each of these three sets, we consider both the global $\Qf$ and the effective $\Qf$ as
described in Section~\ref{sec:plummer}. We show the resulting residuals in
Fig.~\ref{fig:res} where shadow areas show the standard deviation from the models.

There is a remarkable agreement in the predictions between the different analytical
models, no matter if we measure substructure effects or not.  In all cases the accuracy
of the predictions are of order of 10 percentage points.  When carefully measuring the
effects of independent substructure, predictions only improve by 1 percentage point.
Using the effective or the global $Q$ does not improve the estimations significantly.
However, this is because we obtain, in general, high $\fb$ values for simulations in the
turbulent setup, where the difference between global and effective $Q$ is minimal, i.e.,
the mean velocity of the bound entity is similar to the whole cluster when $\fb$ is big
since the bound entity is a considerably large subset of the cluster.   

By eye it appears that there is some improvement from the left panel of
Fig.~\ref{fig:res} to the middle and right panels, however this is not reflected in the
standard errors. 

The results show that the inclusion of an arbitrary distribution for the gas and the
stars does not result in an unpredictable scenario. In fact, the results suggest that it
is possible to estimate how much mass a cluster can retain in any distribution if it is
possible to measure at least the virial ratio of the stellar distribution, without caring
for the presence of substructure or even for the background gas.

\section{Discussion and Conclusions}
\label{sec:conclusion}

We have introduced a simple analytical model that estimates the amount of mass a star
cluster can retain if we instantaneously remove the remaining gas.  We have presented
this model in three flavours: By assuming equal distributions for the gas and the stars,
by carefully measuring the effects of the independent substructure of the gas and stars,
and by only considering the dynamical state of the cluster right after gas is expelled.
We have tested our analytical model by conducting simulations of instantaneous gas
expulsion in highly substructured, embedded star clusters. The amount of substructure
present at the time of gas expulsion was varied in a controlled manner by varying the
treatment of the background gas.  

We find, independent of the treatment of the background gas, the most important
parameters to estimate the survival of a cluster to gas expulsion are the LSF and the
virial ratio at the moment of gas expulsion. However, we also introduce another
independent parameter that works equally well -- \emph{the post-gas-expulsion virial
ratio}. As we are dealing with instantaneous gas expulsion, this is effectively the same
as the pre-gas-expulsion ratio if we only consider the contribution from the stars, and
we disregard the gas contribution altogether. The main advantage of $\Qa$ is that we only
need information about the stars, ignoring completely the presence of the gas.  

The three flavours of the model presented in this paper work equally well in most cases,
estimating final bound fractions with a standard error of $\sim$ 10 percentage points.
The models are not reliable when the bound fractions are low ($\lesssim 40\%$), not
because the models are wrong, but because of technical difficulties when measuring an
accurate virial ratio. An accurate measure of such a value involves information about the
individual members of the cluster that will remain bound after gas expulsion, which is
the information we are trying to estimate in the first place.  Or in other words, it is
more or less impossible to describe the behaviour of a small sub-set of stars by using
global parameters determined using all the stars. 

However, we find it very difficult to obtain low values of $\fb$ when testing our models
in the more ``realistic'' gas distributions.  

Our more ``realistic'' embedded star clusters are obtained from star formation
simulations that arguably are very simplistic but are in agreement with other more
sophisticated simulations: stars form with sub-virial velocities in fractal-like
structures that initially follow the gas distribution, but stars quickly decouple from
the gas during the star cluster formation, trying to form their own independent
distribution.  We find that this mode of star cluster formation is quite stable against
gas removal.  Decoupling from the gas keeps the LSF high and their low stellar velocities
remain low during the star cluster formation process. 

We also follow the evolution of the embedded star cluster after stars form.  We find that
the behaviour of the gas during this phase is critical to ``prepare'' the cluster for gas
expulsion.  We test two extremes of gas evolution: As a first attempt, we stop
gravitational collapse of the gas by switching the equation of state of the gas from
isothermal to adiabatic.  This scenario quickly forms clumps of stars that merges into
bigger clumps and stars couple again with the gas following the overdensities.  As an
alternative approach, we just switch off the self gravity of the gas.  In this case gas
disperses around the cluster and stars and gas remain decoupled since there are no
significant overdensities to follow.  We obtain completely different results in both
cases since, in the first case, the system tends to form a spherical cluster just like in
our previous studies reaching virial equilibrium in the process.  In the second case,
substructure is not erased so efficiently and stars do not have the chance to merge and
virialize in the 2 Myr that we follow their evolution.  Therefore, velocities remain low
at all times in this scenario, and star clusters are able to retain at least $\sim80\%$
of their mass.

While it can be argued that both scenarios are completely unphysical, we wish to note
that reality may be in between.  The equation of the state of the gas in the embedded
phase is not completely understood yet, as it involves complex heating processes.
However, after stars are formed, it is very unlikely that gas is able to accumulate
inside the stellar distribution since radiation from stars would quickly disperse the
interior gas. The scenario (in terms of spatial distribution) may be similar to the
second case when we turned off the gravity of the gas.  Stars will give enough energy to
the gas to support and overcome the gravitational collapse.  If further overdensities
form, they will form more likely outside the star cluster, and therefore will not
contribute to the gravitational potential of the stellar cluster.  The reader should also
keep in mind that all our results are a lower limit of cluster survival.  We use
instantaneous gas expulsion which is the most destructive mode of gas expulsion.  The
timescales of gas expulsion are not known yet, but they cannot be more destructive than
the description used in this study.  

Our results are in close agreement with a similar study realized by
\cite{Kruijssen2012a}, who analyzed the outcome of the hydrodynamical simulations
performed by \citet{Bonnell2003, Bonnell2008} studying the dynamical state of
sub-clusters in the simulations.  They found that stars are formed sub-virial even when
ignoring the background gas and gas fractions inside sub-clusters are small enough to
enable stellar sub-clusters to remain bound when gas is expelled, even in absence of
stellar feedback.  \cite{Lee2016} have also noted that the virial ratio is the only
relevant parameter when estimating bound fractions. Our study compliments this
conclusions by adding that the gas fraction inside the stellar component is not crucial,
as long as the stars are able to remain sub-virial during the embedded phase.  This is
likely to happen if the gas does not form strong overdensities, e.g. is being dispersed,
or when stellar substructure in the star forming region is still important. 

We, therefore, can summarize our main conclusions as follows:
\begin{enumerate}
\item Accurate estimations of the maximum amount of mass that a cluster will loose in the
        transition from the embedded phase to the gas free phase are possible by
        measuring the dynamical state of the stellar component alone ($\Qa$), i.e.,
        ignoring the presence of the gas. 
\item Star clusters formed with low initial velocities are likely to remain in a
        sub-virial state for long time.  If the gas is not able to concentrate and form
        considerable overdensities, then stellar substrucutre lasts longer. 
\item Since erasure of substructure is accompanied by virialization, we find that star
        clusters with high levels of substructure are quite stable against gas expulsion
        no matter how high the gas fraction inside the stellar distribution is.
\end{enumerate}

The first result make estimations on observable embedded star clusters easier since
estimations of the potential energy from molecular clouds are highly challenging.  We
emphasize that it is possible to just ignore the gas to estimate how bound the cluster
is.  Considering that there is theoretical and observable evidence that star clusters
form in sub-virial states \citep[e.g.,][]{Bonnell2003,Bonnell2008} and that feedback
would keep the gas disperse inside the stellar clusters \citep[if present at all,
see][]{Kruijssen2012a} then the conditions in young star clusters are such that they are
very likely to survive gas expulsion, and therefore gas expulsion may not be the culprit
for infant mortality. 

\section*{Acknowledgments}

JPF thanks to  Inti Pelupessy for useful discussions and technical support.  JPF
acknowledge funding to CONICYT through a studentship for Magister students.  MF and RD
acknowledge funding through FONDECYT regular 1130521.  MF also is partly funded by CATA
(BASAL).

\bibliography{bibfile}
\appendix
\section{Resolution criteria for Plummer gas background simulations}\label{sec:resolution}

In SPH, a continuum of gas is transformed into a set of smoothed particles with a certain
smoothing length $h_{i}$. Stars traveling through this sea of particles can be
unphysically deflected when gravitationally interacting with these particles. The
critical velocity at which a star is considerably deflected in our Plummer gas background
setup is
\begin{eqnarray}
        v_{\rm crit} &=& \sqrt{\frac{2GM_{\rm pl}}{N_{\rm gas}h_{i}}}
\end{eqnarray}
\citep[see][]{Hubber2013}, and therefore the velocity dispersion of the stars must be much larger than $v_{\rm
crit}$.

\cite{Hubber2013} developed a resolution criterion for the same experiment we are
exploring, but in a different implementation of the SPH technique.  In this section, we
apply this criterion to the grad-SPH implementation used in this work.

In the current implementation of the SPH technique, $h_{i}$ is obtained by solving the
equation
\begin{eqnarray}
  \frac{4\pi}{3}h_i^3(\rho_i + \rho_{\rm min}) - N_{\rm nb}\langle m_{i}\rangle &=& 0
\end{eqnarray}
\citep{Pelupessy2005}, where $N_{\rm nb}$ is the target number of neighbours, $\rho_{\rm
min}$ is a small density threshold to avoid excessive large $h_i$ at the edges of the
simulation and $\langle m_i \rangle$ is the mean mass of the SPH particles, in our case
this is simply $\langle m_i \rangle = M_{\rm pl}/N_{\rm gas}$.

The smallest value of $h_i$ -- the place with larger $v_{\rm crit}$ -- is the central
region of the Plummer sphere where we have:
\begin{eqnarray}
  \rho_{\rm max} &\approx& \frac{3 M_{\rm pl}}{4\pi R_{\rm pl}^3}.
\end{eqnarray}
The smallest $h_i$ is then
\begin{eqnarray}
h_{\rm min} &\approx&\left( \frac{N_{\rm nb}}{N_{\rm gas}} \right)^{1/3}R_{\rm
pl},
\end{eqnarray}
where we have neglected the contribution of $\rho_{\rm min}$.

Assuming a spherical Plummer-like stellar distribution, the velocity dispersion in the
central region of the gas-star system is
\begin{eqnarray}
  \sigma_{\rm c} &\approx& \sqrt{\frac{1}{6(1-{\rm SFE})}\frac{GM_{pl}}{R_{\rm pl}}},
\end{eqnarray}
where $M_{\rm pl}/(1-{\rm SFE})=M_{\rm tot}$, i.e. the total mass in the cluster. 

Then, the resolution criteria  $\sigma_{\rm c}/v_{\rm crit} \gg 1$ becomes:
\begin{eqnarray}
  \sqrt{\frac{ N_{\rm nb}^{1/3} }{12(1-{\rm SFE} )} } N_{\rm gas}^{1/3} &\gg&1.
\end{eqnarray}

In the Plummer gas background simulations performed in this work we have used $N_{\rm
gas} =100$K, $N_{\rm nb}=64$ and SFE$=0.2$ and the factor $\sigma_{\rm c}/v_{\rm crit} =
30.0$, which is enough to avoid numerical scattering. We note that this factor does not
increase considerably if $N_{\rm gas}$ increases. As shown by \cite{Hubber2011}, a few
thousand particles is enough to accurately reproduce an equilibrium polytrope.

\bsp
\label{lastpage}
\end{document}